%% file: arxiv.tex
\documentclass[letterpaper,10pt]{article}
\usepackage[T1]{fontenc}
\usepackage{parskip}
\usepackage[font={small}]{caption}
\usepackage[hidelinks]{hyperref}
\usepackage[margin=2.5cm]{geometry}
\usepackage{times}
\usepackage{amsmath,amssymb,amsthm}
\usepackage[affil-it]{authblk}
\usepackage{listings}
\usepackage{bm}
\usepackage{xspace}
\usepackage{color}
\usepackage{algorithm,algorithmic,url,listings,enumitem,mathtools}   
\usepackage{graphicx}
\usepackage{subcaption}
\usepackage[round]{natbib}

\usepackage{pgfplots}
\usepackage{grffile}
\pgfplotsset{compat=1.13}
\usetikzlibrary{plotmarks}
\usetikzlibrary{arrows.meta}
\usetikzlibrary{arrows,calc,patterns,decorations.pathmorphing,decorations.markings,positioning}
\usepgfplotslibrary{patchplots}
\newlength\figureheight
\newlength\figurewidth

\newcommand\rv{\xi}

\newcommand{\setX}{\mathsf{X}}
\newcommand{\setY}{\mathsf{Y}}
\newcommand{\reals}{\mathbb{R}}
\newcommand{\myProb}[1]{\mathbb{P}\left(#1\right)}
\newcommand{\myd}{\textrm{d}}      
\newcommand{\ii}{{n}}

\newcommand{\Prb}[1]{\mathbb{P}\left({#1}\right)}
\newcommand{\Exp}[1]{\mathbb{E}\left[{#1}\right]}

\newcommand{\Transp}{\mathsf{T}}
\renewcommand\mid{\,\vert\,}

\newcommand{\T}{T}                                             
\newcommand{\target}{\pi}                                   
\newcommand{\prior}{p}                                       

\newcommand{\range}[2]{#1, \, \dots, \, #2}      

\newcommand{\ConvAS}{\stackrel{\mathrm{a.s.}}{\longrightarrow}}      
\newcommand{\goesto}{\rightarrow}

\newcommand{\Np}{N}                         
\newcommand\uw{w}                 

\newcommand\zhat{\widehat{z}}              

\newcommand{\ssm}{SSM\xspace}

\newcommand{\ie}{i.e.\xspace}

\date{}

\title{Probabilistic learning of nonlinear dynamical systems using sequential Monte Carlo}
\author{Thomas B. Sch\"{o}n\thanks{\url{thomas.schon@it.uu.se}}}
\author{Andreas Svensson}
\author{Lawrence Murray}
\author{Fredrik Lindsten}
\affil{Department of Information Technology, Uppsala University}

\begin{document}

\maketitle

\textbf{Please cite this version:}

Thomas B. Sch\"on, Andreas Svensson, Lawrence Murray and Fredrik Lindsten, 2018. Probabilistic learning of nonlinear dynamical systems using sequential Monte Carlo. In \textit{Mechanical Systems and Signal Processing}, Volume 104, pp. 866-883.

\begin{center}
\begin{minipage}{.75\linewidth}
\begin{lstlisting}[breaklines,basicstyle=\small\ttfamily]
@article{SvenssonSL:2018,
author    = {Sch\"on, Thomas B. and Svensson, Andreas and Murray, Lawrence and Lindsten, Fredrik},
title     = {Probabilistic learning of nonlinear dynamical systems using sequential {Monte} {Carlo}},
journal   = {Mechanical Systems and Signal Processing},
volume    = {104},
year      = {2018},
pages     = {866--883},
}
\end{lstlisting}
\end{minipage}
\end{center}

\vspace{5em}

\begin{abstract}
Probabilistic modeling provides the capability to represent and
manipulate \emph{uncertainty} in data, models, predictions and
decisions. We are concerned with the problem of learning
probabilistic models of dynamical systems from measured
data. Specifically, we consider learning of probabilistic
nonlinear state-space models.  There is no closed-form solution
available for this problem, implying that we are forced to use
approximations.  In this tutorial we will provide a self-contained
introduction to one of the state-of-the-art methods---the particle
Metropolis--Hastings algorithm---which has proven to offer a practical approximation.
This is a Monte Carlo based method, where the particle
filter is used to guide a Markov chain Monte Carlo method through
the parameter space.  One of the key merits of the particle
Metropolis--Hastings algorithm is that it is guaranteed to converge to
the ``true solution'' under mild assumptions, despite being based
on a particle filter with only a finite number of particles. 
We will also provide a motivating numerical example illustrating the method using a modeling language tailored for sequential Monte Carlo methods.
The intention of modeling languages of this kind is to open up the power of sophisticated Monte Carlo methods---including particle Metropolis--Hastings---to a large group of users
without requiring them to know all the underlying mathematical
details.
\end{abstract}

\clearpage


%
\input{introduction}
%
\input{probMod}

%

\input{parInf}
%
\input{UnbiasedEstDL}

%
\input{pmh}

%
\input{varRedPF}
%
\input{numIll}

%
\input{discussion}

\bibliographystyle{abbrvnat}
\bibliography{references}

\clearpage

\appendix
\input{code}

\end{document}

%% file: introduction.tex
\section{Introduction}
%

%
The true value of measured data arises once it has been analyzed and
some kind of knowledge has been extracted from this analysis. The
analysis often relies on the combination of a mathematical model and the
measured data.  The model is a compact representation---set of
assumptions---of some phenomenon of interest, and establishes a link
between that phenomenon and the data, which is expected to provide
some insight. The knowledge we seek is typically a function of some
unknown variables or parameters in the model. However, any reasonable model will be
\emph{uncertain} when making statements about unobserved
variables, and uncertainty therefore plays a
fundamental role in modeling. Probabilistic modeling allows the representation and
manipulation of uncertainty in data, models, decisions and
predictions.  The capability to mathematically represent and
manipulate uncertainty---which is essential to the development
throughout this tutorial---is provided by probability theory. A good
introduction to the ideas underlying probabilistic modeling in
contemporary machine learning is provided by~\cite{Ghahramani:2015}
and from a system identification point of view we
recommend~\cite{Peterka:1981}.

%
Throughout this tutorial we are concerned with the problem of learning
probabilistic models of nonlinear dynamical systems from measured
data, which is sometimes referred to as the nonlinear system
identification problem. These models provide an interpretable
representation from which it is possible to extract the knowledge we seek, compared to doing so directly from the data. The model is thus a natural middle ground between the measured data and that knowledge. More specifically we are concerned with the nonlinear state-space model (SSM)
\begin{subequations}
  \label{eq:SSM}
  \begin{align}
    \label{eq:SSM:dyn}
    x_{t} &= f(x_{t-1}, u_t, v_t, \theta),\\
    \label{eq:SSM:meas}
    y_t &= g(x_t, u_t, \theta) + e_t,\\
    x_0\ &\sim p(x_0 \mid \theta),\\
    \theta &\sim p(\theta),
  \end{align}
\end{subequations}
where $y_t\in\setY$ denotes the observed output and $u_t$ denotes a
known input signal. In the interest of concise notation we will,
without loss of generality, suppress the input signal $u_t$. The
unknown variables are the state $x_t\in\setX$ describing the
system's evolution over time and the static parameters $\theta$. Furthermore, $p(\theta)$ denotes the prior assumptions about $\theta$.
The stochastic variables
$v_t$ and $e_t$ encode noise, commonly referred to as
process noise and measurement noise, respectively. Finally, the
functions $f$ and $g$ encode the dynamics and the measurement equation,
respectively.

The first step towards extracting knowledge from a set of measured
data $y_{1:\T} = \{y_1, \dots, y_{\T}\}$ is to learn a probabilistic
model of the form~\eqref{eq:SSM} by computing the conditional
distribution of the unknown $x_{0:\T}$ and $\theta$ conditioned on
$y_{1:\T}$, denoted $p(x_{0:\T}, \theta \mid y_{1:\T})$. This provides
a useful representation which is typically much closer to the
knowledge we seek than the measured data itself. Once we have a
representation of this conditional distribution, it can be used to
compute more specific quantities that typically constitute the end
result of the analysis. To mention just a few examples of such
quantities we have mean values, variances or estimates of some tail
probability.

The key challenge is that, in general, there is no closed form
expression available for $p(x_{0:\T}, \theta \mid y_{1:\T})$, so
we must resort to approximations. We will focus on approximations based on Monte Carlo sampling which, while being costly to compute, have the appealing property of converging to the true solution as the amount of computations increases. Over the last decade, these
approximations have evolved rapidly, so that we
now have computationally feasible solutions available.

This naturally brings us to the aim of this tutorial, which is to
provide a gentle introduction to probabilistic learning of nonlinear
dynamical systems, and to introduce in some detail one of the current
state-of-the-art methods to do so. This method relies on the
systematic combination of two Monte Carlo algorithms, where a
sequential Monte Carlo algorithm is used to compute a good proposal
distribution for a Markov chain Monte Carlo (MCMC) algorithm.
Hence, in terms of methods, this tutorial is focused on introducing
one particular solution rather than surveying all available methods
(see e.g. \cite{SchonLDWNSD:2015,Kantas:2015} for recent accounts of
that sort).
However, the key ideas discussed in this tutorial (in the context of
the specific method) are in fact central to many other
state-of-the-art Monte Carlo learning methods as well.
For example the particle filter itself, and the fact that it is capable
of producing an unbiased estimate of the likelihood, are also of more
general interest.
In the accompanying paper~\citep{SvenssonLS:2017abc}, we show how
probabilistic modeling, implemented via a new algorithm of the type
outlined in this tutorial, is used to solve one of the challenging
benchmark problems in this special issue with promising results. We will also
hint at the tailored software that is being developed to make these
mathematical tools available to a much wider audience without a
thorough knowledge of Monte Carlo methods. Such software allows the
user to focus entirely on the modeling problem and leave the
computational learning problem to the software.

%
In
Section~\ref{sec:PM} we introduce probabilistic nonlinear
state-space models in more detail. The model as it is stated
in~\eqref{eq:SSM} clearly incorporates uncertainty due to the presence of the
noise sources $v_t$ and $e_t$, as well as uncertainty in the initial
state $x_0$ and in the parameters $\theta$. In probability theory,
uncertainty is represented using random variables and in
Section~\ref{sec:PM} the probabilistic nature of the model will be
made even more explicit when we represent it as a joint distribution
$p(x_{0:\T}, \theta, y_{1:\T})$ of all the random variables present in the model. The basic
Monte Carlo idea is then introduced in
Section~\ref{sec:ParameterInference} together with an explanation of
how this idea can be used to learn the conditional distribution of the
parameters given the measurements $p(\theta\mid y_{1:\T})$. The idea
is developed further in Section~\ref{sec:SE}, where it becomes clear
that we also need information about the unknown state variables,
resulting in the introduction of the sequential Monte Carlo method
(a.k.a. the particle filter) to estimate the state variables. In Section~\ref{sec:PMH}, the particle filter is used inside the MCMC
method introduced in Section~\ref{sec:ParameterInference}. The basic
particle filter construction from Section~\ref{sec:SE} can be improved
in several ways and in Section~\ref{sec:VarRedPF} we discuss some of
the most important developments in this direction. The resulting
method is then illustrated using a nonlinear spring-damper system in
Section~\ref{sec:NumIll}. Finally we conclude with a discussion in
Section~\ref{sec:discussion}.


%% file: probMod.tex
\section{Probabilistic modelling of dynamical systems}
\label{sec:PM}

We will refer to the joint distribution of all observed (here $y_{1:\T}$) and unobserved (here $x_{0:\T}$ and~$\theta$) variables as the \emph{full probabilistic model}, which in our present setting amounts to $p(x_{0:\T}, \theta, y_{1:\T})$. The idea of using the mathematics of probability to represent and manipulate uncertainty is commonly referred to as Bayesian statistics \citep{GelmanCSDVR:2013}. In order to write down the full probabilistic model for~\eqref{eq:SSM} let us start by noticing that for many models we can express the conditional distribution of $y_{t}$ given $x_t$ and $\theta$ as
\begin{align}
	\label{eq:PM:py}
  p(y_{t}\mid x_t, \theta) = p_{e_t}(y_{t} - g(x_t, \theta), \theta),
\end{align}
where $p_{e_t}(\cdot)$ denotes the distribution of the measurement noise $e_t$. Looking back at~\eqref{eq:SSM:meas} with the assumption that $x_t$ and $\theta$ are both known we see that the only randomness stems from the measurement noise $e_t$. Hence, the conditional distribution of $y_{t}$ given $x_t$ and $\theta$ is dictated by $p_{e_t}(\cdot)$, which explains~\eqref{eq:PM:py}. In a similar way, the dynamical equation \eqref{eq:SSM:dyn} implicitly defines a conditional probability distribution $p(x_{t} \mid x_{t-1}, \theta)$ describing the state evolution.\footnote{The attentive reader might wonder why the noise is assumed to be additive in~\eqref{eq:SSM:meas} but not in \eqref{eq:SSM:dyn}. The reason is that the methods that we will consider require the density function $p(y_t \mid x_t, \theta)$ to be available for point-wise evaluation. The additivity assumption implies that this density can be expressed as in \eqref{eq:PM:py}. The methods do not, however, require $p(x_{t} \mid x_{t-1}, \theta)$ to be available for point-wise evaluation.} Hence, the \ssm in~\eqref{eq:SSM} can be expressed as
\begin{subequations}
	\label{eq:SSMpdf}
	\begin{align}
    	x_{t}\mid x_{t-1},\theta &\sim p(x_{t}\mid x_{t-1}, \theta),\\
	    y_t\mid x_t,\theta &\sim p(y_t\mid x_t, \theta),\\
    	\theta &\sim p(\theta)
	\end{align}
\end{subequations}
An important and useful aspect of the \ssm is that the current state $x_t$ contains all information about the past that is needed to make predictions into the future. Formally this aspect is captured by the \emph{Markov property} stating that $p(x_{t+1}\mid x_{0:t}) = p(x_{t+1}\mid x_{t})$. Using conditional probabilities we can factor the full probabilistic model as
\begin{align}
  \label{eq:dd1}
  p(x_{0:\T}, \theta, y_{1:\T}) = p(y_{1:\T}\mid x_{0:\T},
  \theta)
  \underbrace{p(x_{0:\T}\mid \theta)p(\theta)}_{\text{prior distribution}},
\end{align}
where $p(y_{1:\T}\mid x_{0:\T}, \theta)$ describes the distribution of the data. The so-called \emph{prior distribution} $p(x_{0:\T}, \theta) = p(x_{0:\T} \mid \theta)p(\theta)$ represents our initial assumptions about the unknown states and parameters. It is instructive to rewrite the model~\eqref{eq:dd1} slightly in order to more clearly see that it is just a different way of representing~\eqref{eq:SSM}. Let us first continue the use of conditional probabilities in order to further decompose $p(y_{1:\T}\mid x_{0:\T}, \theta)$ into
\begin{align}
  \notag
  p(y_{1:\T}\mid x_{0:\T}, \theta) &= p(y_{\T} \mid
   y_{1:\T-1},x_{0:\T}, \theta) p(y_{1:\T-1} \mid x_{0:\T}, \theta)\\
  \label{eq:dataDist}
   &= p(y_{\T} \mid x_{\T}, \theta) p(y_{1:\T-1} \mid x_{0:\T-1}, \theta),
\end{align}
where we also made use of the conditional independence of the observations given the current state, as well as the Markov property of the state. Moreover we can rewrite the prior distribution over the states $p(x_{0:\T}\mid \theta)$ by noting that
\begin{align}
  \label{eq:StatePrior}
  p(x_{0:\T}\mid \theta) = p(x_{\T} \mid x_{0:\T-1}, \theta)
  p(x_{0:\T-1} \mid \theta) =  p(x_{\T} \mid x_{\T-1}, \theta) p(x_{0:\T-1} \mid \theta),
\end{align}
where we made use of conditional probabilities in the first equality and the Markov property in the second equality. Repeated use of~\eqref{eq:dataDist} and~\eqref{eq:StatePrior} results in
\begin{align}
  \label{eq:SSM2}
  p(x_{0:\T}, \theta, y_{1:\T}) = \left(\prod_{t=1}^{\T}\underbrace{p(y_t\mid x_t,
  \theta)}_{\text{observation}} \right)
  \underbrace{\left(\prod_{t=1}^{\T} \underbrace{p(x_{t}\mid x_{t-1}, \theta)}_{\text{dynamics}} \right)
  \underbrace{p(x_0\mid \theta)}_{\text{initial}}\underbrace{p(\theta)}_{\text{param.}}}_{\text{prior}}
\end{align}
explicitly showing how the ``engineering standard'' \ssm, as it is formulated in~\eqref{eq:SSM}, relates to the full probabilistic model formulation.

Starting from our model~\eqref{eq:SSM2} (or~\eqref{eq:SSM}), the problem we set out to solve is how to compute the distribution of the unobserved variables conditioned on the observed variables, $p(x_{0:\T}, \theta\mid y_{1:\T})$. This distribution is referred to as the \emph{posterior} distribution. Using conditional probability it separates into the so-called state and parameter inference problems according to
\begin{align}
  \label{eq:PostDist}
  p(x_{0:\T}, \theta\mid y_{1:\T}) = \underbrace{p(x_{0:\T} \mid \theta,
  y_{1:\T})}_{\text{state inf.}} \underbrace{p(\theta \mid
  y_{1:\T})}_{\text{param. inf.}}.
\end{align}
In solving the parameter inference problem---which is the focus of this
tutorial---it is helpful to start with
\begin{align}
  \label{eq:targetPE}
  p(\theta\mid y_{1:\T}) = \frac{p(y_{1:\T}\mid
  \theta)p(\theta)}{p(y_{1:\T})} = \frac{p(y_{1:\T}\mid
  \theta)p(\theta)}{\int p(y_{1:\T}\mid \theta)p(\theta)\myd\theta}.
\end{align}
The target for the inference algorithms we eventually will derive will mainly be $p(\theta\mid y_{1:\T})$,
and for this reason we refer to it as the \emph{target distribution} and denote it by $\target(\cdot)$.
The end user may not be
interested in the target distribution per se, but rather in
some test function $\varphi(\theta)$ evaluated over it. These can be
evaluated by solving integrals of the form
\begin{align}
  \label{eq:PI:TestFcn}
	I[\varphi] =   \Exp{\varphi(\theta)\mid y_{1:\T}} = \int \varphi(\theta)p(\theta\mid y_{1:\T}) \myd\theta.
\end{align}

A common test function is just the identity function
$\varphi(\theta) = \theta$, so that the integral~\eqref{eq:PI:TestFcn}
provides an estimate of the expected value of~$\theta$ conditioned on
the data~$y_{1:\T}$, \ie $\Exp{\theta\mid y_{1:\T}}$. Another example
is the indicator function $\varphi(\theta) = I(\theta > \vartheta)$,
for some threshold value $\vartheta$, which provides an estimate of a
tail probability, perhaps important in modelling extreme events. Other
test functions or combinations of them yield the covariance and higher
moments, or estimates of domain-specific utility or loss.

From~\eqref{eq:targetPE}, it is clear that in order to compute the posterior distribution of the unknown  parameters we first need to compute the \emph{data distribution} $p(y_{1:\T}\mid \theta)$. This can be expressed in terms of the joint distribution $p(x_{0:\T}, y_{1:\T}\mid \theta)$ by averaging it (marginalizing) over all possible values for the state variables $x_{0:\T}$
\begin{align}
  \label{eq:DataDist}
  p(y_{1:\T}\mid \theta) = \int p(y_{1:\T}, x_{0:\T} \mid \theta) \myd x_{0:\T}.
\end{align}
Besides being useful to us later, this indicates the important fact that the state inference problem is inherent in the parameter inference problem when we are dealing with \ssm{s}. Note that if we instead model the unknown parameters~$\theta$ as deterministic variables to be estimated using maximum likelihood, we still have to deal with the high-dimensional integral in~\eqref{eq:DataDist}, since this will be used in computing the so-called likelihood function. The likelihood function is defined as the data distribution evaluated for the particular measurement sequence $y_{1:\T}$ that we have available, and it is thus not a probability distribution, but rather a deterministic function of the unknown variables $\theta$. Hence, both Bayesian and maximum likelihood approaches will benefit from the methodology presented in this paper.

We will, however, reuse an important term from the maximum likelihood literature, and simply refer to the value of the data distribution $p(y_{1:\T}\mid \theta)$ evaluated at $y_{1:T}$ and $\theta$ as \emph{the likelihood}. Furthermore, we have used---and will continue to use---$p(\cdot)$ to denote a probability density function of variables that in some way describes the model or can be induced from the model~\eqref{eq:SSM}. Distributions describing random variables that are not directly related to the model are instead denoted using characters from the Greek alphabet. These variables will for example arise as parts of the (stochastic) algorithms that we will derive in the coming sections.

%% file: parInf.tex
\section{Solving the parameter inference problem}
\label{sec:ParameterInference}
The parameter inference problem amounts to computing the posterior distribution $p(\theta\mid y_{1:\T})$, and commonly also to evaluate the integral~\eqref{eq:PI:TestFcn} with respect to this distribution for some test function $\varphi(\theta)$. The posterior $p(\theta\mid y_{1:\T})$ is in general\footnote{For certain special cases such as the autoregressive model with only Gaussian noise, closed-form solutions exist. This corresponds to Bayesian linear regression \citep{GelmanCSDVR:2013,Bishop:2006}.} not available in closed form, and consequently there are typically no closed-form solutions to the integral~\eqref{eq:PI:TestFcn} either. 
We can, however, construct an estimator of it using the Monte Carlo idea,
which is introduced in Section~\ref{sec:MCidea}. More specifically we
will make use of a so-called pseudo-marginal 
MCMC 
method--- introduced in Section~\ref{sec:pmMCMC}---which itself
employs an estimate. However, before we describe the pseudo-marginal
idea we will first introduce the basic MCMC idea itself in
Section~\ref{sec:MCMC}.

\subsection{The Monte Carlo idea}
\label{sec:MCidea}
Monte Carlo integration provides approximate solutions to integrals of
the form
\begin{align}
  \label{eq:int}
  c = \Exp{h(\rv)} \triangleq \int_\mathsf{\Xi}h(\rv)p_\rv (\rv)\myd \rv,
\end{align}
where $\rv\in\mathsf{\Xi}$ and $\rv\sim p_\rv(\rv)$. A classical
choice to evaluate the integral is to take $N$ equally-spaced grid
points $\{{\rv}^\ii\}_{\ii=1}^N$ on some interval $[a,b]$, and compute (for the scalar case)
\begin{align}
  \label{eq:RiemannSum}
  \widehat{c} = \frac{b-a}{N}\sum_{\ii=1}^{N}h(\rv^\ii)p(\rv^\ii)
\end{align}
as an estimate of $c$. This is the usual Riemann sum. In Monte Carlo
integration, the idea is to instead choose $N$ \emph{random} sample points
$\{{\rv}^\ii\}_{\ii=1}^N$ from the probability distribution $p_\rv (\rv)$ and
compute
\begin{align}
  \label{eq:MCint}
  \widehat{c} = \frac{1}{N}\sum_{\ii=1}^{N}h(\rv^\ii).
\end{align}
Clearly $\widehat{c}$ is now itself random. There are two advantages
to this approach. The first is that, while a fixed grid can give an
arbitrarily bad estimate, it can be proven that the expectation of the
random estimator $\widehat{c}$ from the Monte Carlo integration~\eqref{eq:MCint} is exactly
$c$, i.e., $\Exp{\widehat{c}} = c$, and we say that $\widehat{c}$ is
an \emph{unbiased} estimator of $c$. Secondly, if
the dimension of $\mathsf{\Xi}$ is $D$, the error of the Riemann
sum~\eqref{eq:RiemannSum} scales as $\mathcal{O}(N^{-1/D})$, while
that of the Monte Carlo estimate~\eqref{eq:MCint} scales as
$\mathcal{O}(N^{-1/2})$ \citep[see
e.g.][]{Robert:2004,Owen:2013}. That is, for more than a few
dimensions, the Monte Carlo estimator exhibits smaller error.

\subsection{The Markov chain Monte Carlo idea}
\label{sec:MCMC}
A Markov chain $\{\theta[m]\}_{m\geq 0}$ is a stochastic process with
the property that the next state $\theta[m+1]$ of the process depends
on the current state $\theta[m]$, but, given this, it is conditionally independent
of all previous states $\theta[{0:m-1}]$, recall the Markov property introduced in Section~\ref{sec:PM}. It is completely specified by
an initial distribution and a transition kernel which defines the (stochastic) transition from $\theta[m]$ to $\theta[m+1]$.  The
SSM~\eqref{eq:SSM} is a popular and useful example of a Markov chain
$\{x_t\}_{t\geq 0}$ with initial distribution $p(x_0)$ and transition
kernel given by $p(x_{t}\mid x_{t-1})$.
A Markov chain is said to have an \emph{equilibrium distribution} if the marginal distribution of the samples generated by the chain, $\{\theta[m]\}_{m=0}^{M}$, converges to this equilibrium distribution asymptotically as  $M\to\infty$. That is, for large $m$, the samples $\theta[m]$ will be distributed roughly to the equilibrium distribution regardless of the distribution of $\theta[0]$.
The idea underlying MCMC is to simulate a Markov chain which is designed in such a way that its equilibrium distribution coincides with the target distribution, here taken to be $\target(\theta) = p(\theta\mid y_{1:\T})$. Our interest lies in Markov chains which are structured such that as $M \goesto \infty$ we have that
\begin{align}
  \label{eq:MCMCergodicThm}
  \frac{1}{M}\sum_{m=0}^{M}\varphi(\theta[m])  \ConvAS \int \varphi(\theta) p(\theta \mid y_{1:\T}) \myd\theta,
\end{align}
where $\ConvAS$ denotes almost sure convergence.
The finite and explicit sum in~\eqref{eq:MCMCergodicThm} provides an approximation of the intractable integrals \emph{if} there is a way of constructing the underlying Markov chain $\{\theta[m]\}_{m\geq 0}$. The construction of such a Markov chain is indeed possible and one way is to use the following two-step procedure. 
In the first step a new
\emph{candidate} state $\theta^{\prime}$ is proposed by generating a
sample from the so-called \emph{proposal distribution}
$q(\theta\mid \theta[m])$, i.e.
$\theta^{\prime} \sim q(\theta\mid \theta[m])$. The proposal
distribution depends on the current state $\theta[m]$ and its
form is decided by the user. A common choice is a so-called Gaussian random walk, where a candidate sample is obtained by simply adding a realization from a certain Gaussian random variable to the current state of the Markov chain. 

In the second step we must choose whether the next state $\theta[m+1]$
of the Markov chain should be the candidate sample $\theta^{\prime}$
just proposed, or a repeat of the current state
$\theta[m]$. It can be shown that if we chose the candidate sample as
the next state with probability
\begin{align}
  \label{eq:MHAP}
  \alpha =
  \min\left(1,\frac{\target(\theta^\prime)}{\target(\theta[m])}\frac{q(\theta[m]\mid
  \theta^\prime)}{q(\theta^\prime\mid \theta[m])}\right)
  = \min\left(1,\frac{p(y_{1:\T}\mid \theta^{\prime})\prior(\theta^{\prime})}{p(y_{1:\T}\mid \theta[m])\prior(\theta[m])}
  \frac{q(\theta[m]\mid \theta^\prime)}{q(\theta^\prime\mid \theta[m])}\right),
\end{align}
the resulting Markov chain will have the target distribution as its
equilibrium distribution. The second equality in~\eqref{eq:MHAP} is due
to~\eqref{eq:targetPE}. For natural reasons $\alpha$
in~\eqref{eq:MHAP} is referred to as the \emph{acceptance
  probability}; with probability $\alpha$ the new state in the
Markov chain is chosen as $\theta[m+1] = \theta^{\prime}$ (the
candidate sample is accepted) and with probability $1- \alpha$ the
candidate sample is rejected and the Markov chain remains in the same
state as in the previous iteration, i.e.  $\theta[m+1] = \theta[m]$.
The samples $\{\theta[m]\}_{m=0}^{M}$ from the resulting Markov chain
constitutes an empirical approximation
\begin{align}
  \label{eq:MHtargetEst}
  \widehat{\target}(\theta) = \frac{1}{M}\sum_{m=0}^{M} \delta_{\theta[m]}(\theta)
\end{align}
of the posterior distribution. Here, $\delta_{\theta[m]}(\theta)$
denotes the (Dirac) point-mass distribution at $\theta = \theta[m]$.
The algorithm of first proposing a candidate state and then accepting
or rejecting this as the next state of the Markov chain is referred to
as the \emph{Metropolis--Hastings (MH) algorithm}, introduced
by \cite{Metropolis:1953,Hastings:1970}.  By now there are many
textbook style introductions available, see
e.g. \cite{Robert:2004,Owen:2013}.

The intractable integral~\eqref{eq:PI:TestFcn} becomes a
tractable finite sum simply by inserting~\eqref{eq:MHtargetEst}
into~\eqref{eq:PI:TestFcn}, resulting in
\begin{align}
  \widehat{I}[\varphi] \triangleq \int \varphi(\theta)
  \widehat{\target}(\theta) \myd \theta
  = \frac{1}{M}\sum_{m=0}^{M}\int \varphi(\theta)
  \delta_{\theta[m]}(\theta) \myd \theta
  = \frac{1}{M}\sum_{m=0}^{M} \varphi(\theta[m]),
\end{align}
where the last equality follows from the properties of the Dirac point-mass distribution. The estimator $\widehat{I}[\varphi]$ defined
above is well-behaved in the sense that under certain non-trivial conditions it obeys both a law of large
numbers and a central limit theorem, see e.g. \cite{Robert:2004,MeynT:2009} for a thorough treatment. The law of
large numbers tells us that the estimator is consistent as $M\goesto \infty$, whereas the central
limit theorem tells us that the error is approximately Gaussian with a
standard deviation that decreases with the typical Monte Carlo rate of $1/\sqrt{M}$.  A
nice historical account of MCMC is provided in \cite{RobertC:2011}.

\subsection{Using unbiased estimates within Metropolis--Hastings}
\label{sec:pmMCMC}
The problem preventing us from implementing the Metropolis--Hastings
algorithm to infer $p(\theta\mid y_{1:T})$ for a general state-space model is that we cannot compute
the acceptance probability $\alpha$ given in~\eqref{eq:MHAP}, since
there is no closed-form expression available for the likelihood
$p(y_{1:\T}\mid \theta)$. However, what if we have an estimate of the
likelihood, can we then use this estimate instead of the exact
likelihood in the acceptance probability and still
end up with a valid algorithm? (``Valid'', here, meaning that the
method converges in the sense of \eqref{eq:MCMCergodicThm}.) The
answer to this highly nontrivial question is actually
\emph{yes}---under certain conditions on the likelihood estimate
(described below).
%
%
The result is referred to as an \emph{exact approximation}, since we
obtain an \emph{exact} Metropolis--Hastings algorithm, in the sense
that the equilibrium distribution of the Markov chain remains the target distribution of interest, despite the fact that we make use
of an approximation of the likelihood when evaluating the
acceptance probability.

Let us now explain and prove that this actually works. We start by assuming that we have access to an estimator $\zhat$ of the likelihood $p(y_{1:T}\mid \theta)$. The estimator naturally depends on
the observed data $y_{1:T}$ and the model parameter $\theta$ (since
these variables determine the value of the estimand). Furthermore, we
assume that the estimator depends on some additional random
variables---as we shall see below, these random variables typically
come from yet another Monte Carlo procedure which is used to compute
the estimate. Consequently, the estimator $\zhat$ is itself a \emph{random variable} and it has some distribution $\psi(z \mid \theta, y_{1:T})$ depending on $\theta$ and $y_{1:T}$. Furthermore, we will not require $\psi$ to be available on closed form (as we will see its density will actually cancel in the acceptance rate, so we do not need to evaluate it), but it exists conceptually nevertheless.

Next, we introduce the random variable $\zhat$ (\ie, the estimator of the likelihood) as an \emph{auxiliary variable} in the model. Auxiliary variables---while being of no particular interest on their own---are commonly introduced in statistical models in order to simplify the inference for some other variable of interest. In our case, the variable of interest is the model parameter $\theta$. Thus, we consider the joint distribution for $(\theta,z)$ given by
\begin{align}
	\label{eq:PMMH:ext-target-wontwork}
  \psi(\theta, z\mid y_{1:\T}) =
  p(\theta \mid y_{1:T}) \psi(z \mid \theta, y_{1:T}) =
  \frac{p(y_{1:\T}\mid \theta) \prior(\theta) \psi(z\mid\theta,y_{1:T})}{p(y_{1:\T})}.
\end{align}
We note that the original target distribution $p(\theta\mid y_{1:\T})$ is (by construction) obtained by marginalizing the joint distribution $\psi(\theta, z\mid y_{1:\T})$ with respect to the auxiliary variable~$\zhat$.

If we now were to construct a Metropolis--Hastings algorithm for both
the parameters $\theta$ and the auxiliary variable $\zhat$ it would suffer from the
same problem as before, since the (intractable) likelihood still
appears in the expression~\eqref{eq:PMMH:ext-target-wontwork}.
To overcome this difficulty we will make use of the following trick: we \emph{define} a new joint target distribution over $(\theta,\zhat)$ by simply replacing the intractable likelihood $p(y_{1:T}\mid \theta)$ in \eqref{eq:PMMH:ext-target-wontwork} with its estimator~$\zhat$
\begin{align}
  \label{eq:PMMH:ext-target-willwork}
	\target(\theta, z\mid y_{1:\T}) := \frac{z \,\prior(\theta) \psi(z\mid\theta,y_{1:T})}{p(y_{1:\T})}.
\end{align}
For this distribution to be useful for our purposes we have three requirements that need to be fulfilled. The first two are formal in nature (stating that $\target(\theta, z\mid y_{1:\T})$ is a valid probability distribution): \emph{(i)} that $\target(\theta, z\mid y_{1:\T})$ has to be everywhere non-negative, and \emph{(ii)} that it integrates to 1. The third requirement stems from the auxiliary variable construction, stating \emph{(iii)} that $\target(\theta, z\mid y_{1:\T})$ has to preserve the property that its marginal distribution for $\theta$ coincides with the target distribution of interest: $\int \target(\theta, z \mid y_{1:T}) dz = p(\theta \mid y_{1:T})$. As we shall see next, all of these requirements are fulfilled if $\zhat$ is a \emph{non-negative} and \emph{unbiased} estimate of $p(y_{1:T} \mid \theta)$.

The first requirement---non-negativity of $\target$---follows directly from the assumed non-negativity of $\zhat$. For the second and third requirements, we consider the integral
\begin{align}
	\int \target(\theta, z \mid y_{1:T}) \myd z = \frac{\prior(\theta)}{p(y_{1:\T})} \int z\, \psi(z\mid\theta,y_{1:t}) \myd z.
\end{align}
However, since $\psi(z\mid\theta,y_{1:T})$ is nothing but the distribution of the estimator $\zhat$, the integral in the expression above is just the mean of $\zhat$. Hence, due to the assumed unbiasedness of $\zhat$
\begin{align}
	\label{eq:PMMH:correct-marginal-recovered}
	\int \target(\theta, z \mid y_{1:T}) \myd z
	= \frac{\prior(\theta)}{p(y_{1:\T})} p(y_{1:T} \mid \theta)  = p(\theta \mid y_{1:T}).
\end{align}
Thus, we see that the marginal distribution of $\target$ with respect to $\theta$ is $p(\theta \mid y_{1:T})$ as required, which also implies that $\target$ is a properly normalized probability distribution (it integrates to 1, which can be seen by further integrating both sides of \eqref{eq:PMMH:correct-marginal-recovered} with respect to $\theta$).

We will now construct a standard Metropolis--Hastings algorithm with $\target(\theta, z \mid y_{1:T})$ as its target distribution. Note that we will work on the joint space of $\theta$ and $\zhat$, and the joint proposal will be a two-step procedure which first samples $\theta^{\prime}$ from $q(\theta\mid\theta[m])$, and thereafter samples $\zhat^{\prime}$ from $\psi(z\mid\theta^{\prime}, y_{1:T})$. More compactly, we can write this as $(\theta^{\prime}, \zhat^{\prime}) \sim  q(\theta^{\prime}\mid\theta[m])\psi(z^{\prime}\mid\theta^{\prime}, y_{1:T})$. Inserting all these expressions into the acceptance rate~\eqref{eq:MHAP}, we obtain
\begin{align}
	\notag
	\alpha &=
	\min\left(1,\frac{\target(\theta', z' \mid y_{1:T})}{\target(\theta[m], z[m] \mid y_{1:T})}\frac{q(\theta[m]\mid\theta')\psi(z[m]\mid\theta[m], y_{1:T})}{q(\theta'\mid\theta[m])\psi(z'\mid\theta', y_{1:T})}\right) \\
	\notag
	&=
	\min\left(1,\frac{z'p(\theta')\psi(z' \mid \theta', y_{1:T})}{z[m]p(\theta[m])\psi(z[m] \mid \theta[m], y_{1:T})}
	\frac{q(\theta[m]\mid\theta')\psi(z[m]\mid\theta[m], y_{1:T})}{q(\theta'\mid\theta[m])\psi(z'\mid\theta', y_{1:T})}\right) \\	
	\label{eq:PMMHAP}
	&= \min\left(1,\frac{z^{\prime}\prior(\theta^{\prime})}{z[m]\prior(\theta[m])}
	\frac{q(\theta[m]\mid \theta^\prime)}{q(\theta^\prime\mid \theta[m])}\right).
\end{align}
The development above is summarize in Algorithm~\ref{alg:pseudoMargMP}.

\begin{algorithm}[ht]
	\caption{\textsf{pseudo-marginal Metropolis--Hastings}}
	\small
	\begin{algorithmic}[1]
		\STATE \textbf{Initialisation ($m=0$):} Set $\theta[0]$ arbitrarily and sample $\zhat[0] \sim \psi(z \mid \theta [0], y_{1:\T})$.
		\FOR{$m = 1$ \textbf{to} $M$}
		\STATE Sample $\theta^{\prime} \sim q(\theta \mid \theta[m-1])$.
		\STATE Sample $\zhat^{\prime} \sim \psi(z \mid \theta^{\prime}, y_{1:\T})$.
		\STATE With probability
		\begin{align}
		\alpha = \min\left(1,\frac{\zhat^{\prime}\prior(\theta^{\prime})}{\zhat[m-1]\prior(\theta[m-1])}
		\frac{q(\theta[m-1]\mid \theta^\prime)}{q(\theta^\prime\mid \theta[m-1])}\right),
		\end{align}
		set $\{\theta[m], \zhat[m]\} \leftarrow \{\theta^{\prime}, \zhat^{\prime}\}$
		(accept the candidate samples) and with probability $1-\alpha$ set
		$\{\theta[m], \zhat[m]\} \leftarrow {\{\theta[m-1], \zhat[m-1]\}}$ (reject the
		candidate samples).
		\ENDFOR
	\end{algorithmic}
	\label{alg:pseudoMargMP}
\end{algorithm}

This idea of making use of a non-negative and unbiased estimate of the likelihood within a Metropolis--Hastings algorithm is referred to as the \emph{pseudo-marginal approach}. It was first introduced in $2003$~\cite{Beaumont:2003} and it has later been generalized and thoroughly analysed, see~\cite{AndrieuR:2009,AndrieuV:2015}. Note that it is only required that we can simulate from $\psi(z\mid \theta, y_{1:\T})$, we never have to evaluate it point-wise.

The pseudo-marginal algorithm will converge towards the correct posterior distribution $p(\theta \mid y_{1:T})$ under weak assumptions, as long as the nonnegativity and unbiasedness conditions hold for the estimate $\zhat$ (convergence, here, means that the distribution of $\theta[m]$ converges to $p(\theta \mid y_{1:T})$ as $m \goesto \infty$). This is not to say that the precision of the estimate $\zhat$ is of no interest, however. Indeed, if the variance in the estimate $\zhat$ is overly large, this will result in the algorithm converging slowly. In practice it is therefore important to keep the variance in the estimate $\zhat$ as low as possible to obtain an efficient solution. We return to this in Section~\ref{sec:PMH}.

%% file: UnbiasedEstDL.tex
\section{Computing an unbiased estimate of the likelihood using the particle filter}
\label{sec:SE}

In order to use Algorithm~\ref{alg:pseudoMargMP} for identification of a state-space model, we need a way to construct an \emph{unbiased} and \emph{non-negative} estimate of the likelihood $p(y_{1:T}\mid\theta)$ for any value of $\theta$.  Specifically, as the algorithm considers a candidate value $\theta'$ for the model parameters, we need to compute an estimate $\zhat \approx p(y_{1:T} \mid \theta')$ that is, intuitively speaking, used to determine whether or not $\theta'$ is a promising candidate.

The likelihood is only a function of the data $y_{1:T}$ and the parameters $\theta$, and not the state variables $x_{0:T}$. Thus, to compute $p(y_{1:T}\mid\theta)$ we need to marginalize out all possible trajectories $x_{0:T}$. We can write this as
\begin{align}
 \label{eq:dlik}
  p(y_{1:T}\mid\theta) = \int_{\mathsf{X}^{T+1}} p(y_{1:T}\mid
                       x_{0:T},\theta) p(x_{0:T}\mid \theta) \myd x_{0:T}.
\end{align}
In general, this integral admits no closed-form solution\footnote{For
  certain (indeed important) special cases such as linear models with
  only Gaussian noise, closed-form solutions exists, obtained via the
  Kalman filter. Complete details on that special case---including how to use Kalman filters to implement a Metropolis--Hastings algorithm---is provided in~\cite{SchonLDWNSD:2015}.}. We can, however, construct an unbiased and
non-negative estimator of it using the Monte Carlo idea. We will in Section~\ref{sec:MCintfp} detail a naive construction of such an estimator. That construction serves as a motivation for the particle filter, which is then introduced in Section~\ref{sec:PF}. It provides a practically useful computation of a non-negative and unbiased estimator of the likelihood for a nonlinear state-space model~\eqref{eq:SSM}.

\subsection{Monte Carlo integration for the likelihood}
\label{sec:MCintfp}
The problem of solving~\eqref{eq:dlik} is a special case of~\eqref{eq:int} where $\rv = x_{0:T}$, $p_\rv(\rv)
= p(x_{0:T}\mid \theta)$ and $h(\rv) = p(y_{1:T}\mid x_{0:T},\theta)$. A
first attempt to solve this problem is to use the basic Monte Carlo approach as presented in Section~\ref{sec:MCidea} and proceed as follows:
\begin{enumerate}
\item Generate $N$ samples of the state trajectory $x_{0:T}^\ii \sim p(x_{0:T} \mid \theta)$ for $\ii = \range{1}{N}$. In practice, this can be done by  simulating the system dynamics: $x_0^\ii \sim \prior(x_0 \mid \theta)$, then $x_1^\ii \sim p(x_1 \mid x_0^\ii, \theta)$, then $x_2^\ii \sim p(x_2 \mid x_1^\ii, \theta)$, etc. up to $T$, for $\ii = \range{1}{N}$.
\item Compute an estimate of the likelihood as $\zhat := \frac{1}{N}\sum_{\ii=1}^{N} p(y_{1:T} \mid x_{0:T}^\ii,\theta) = \frac{1}{N}\sum_{\ii=1}^{N} \prod_{t=1}^T p(y_t \mid x_t^\ii, \theta)$.
\end{enumerate}
This results in a basic Monte Carlo estimate $\zhat$ which is indeed unbiased and non-negative so it is a valid approach. In practice, however, the variance of this $\zhat$ is too large to be of any practical use, as we will illustrate in Figure~\ref{fig:likelihoodest}. The reason is that many samples $\{x_{0:T}^\ii\}_{\ii=1}^N$ are likely to be drawn in a place where they contribute very little to the estimate due to the high dimension of the space $\mathsf{X}^{(T+1)}$.

To mitigate this problem the structure of the state-space model can be leveraged.  Specifically, we factorise the high-dimensional integral into a product of $\T$ lower-dimensional integrals:
\begin{align}
  \label{eq:dlikfactor}
  p(y_{1:T} \mid \theta) = \prod_{t=1}^{T} p(y_t\mid y_{1:t-1},\theta)
  = \prod_{t=1}^{T} \int_\mathsf{X} p(y_t\mid x_t, \theta) p(x_t \mid
  y_{1:t-1},\theta) \myd x_t.
\end{align}
We can then perform Monte Carlo integration for each integral, one at a time, and take the product of the estimates to obtain an overall estimate. The advantage here is that it is typically much easier to handle these lower-dimensional integrals than it is to handle the preceding higher-dimensional integral. The drawback, however, is that it is typically not possible to simulate directly from $p(x_{t} \mid y_{1:t-1},\theta)$. To tackle this issue we will instead make use of an algorithm known as the \emph{particle filter}, which we introduce next.

\subsection{Introducing the particle filter}\label{sec:PF}%
The particle filter sequentially generates samples $\{x_t^\ii\}_{\ii=1}^N$ for $t=1,\,2,\,\dots$, such that the samples with index $t$ are approximately distributed according to $p(x_t \mid y_{1:t-1}, \theta)$. What differs from the basic Monte Carlo procedure outlined above (where each $x_{0:T}^n$ was drawn independently) is that all the samples are allowed to interact between iterations.
Note that the particle filter is most often derived as a way of approximating the filter distribution $p(x_t \mid y_{1:t}, \theta)$. However, we can equally-well derive it as an algorithm for approximation of the predictor distribution $p(x_t \mid y_{1:t-1}, \theta)$, since that is what we need to estimate the likelihood.
To derive the particle filter it is convenient to introduce the notion of an \emph{empirical approximation} of a probability distribution. An empirical distribution with $N$ samples $\{x_t^\ii\}_{\ii=1}^N$ approximating the distribution $p(x_t \mid y_{1:t-1}, \theta)$ looks as
\begin{align}
 \label{eq:PF:EmpiricalApprox}
  \widehat{p}(x_t \mid y_{1:t-1},\theta) = \frac{1}{N}\sum_{\ii=1}^N \delta_{x^\ii_t}(x_t),
\end{align}
where $\delta_{x^\ii_t}(x_t)$ is again a (Dirac) point-mass distribution at $x_t^\ii$.  Note that if we plug this empirical distribution into an integral with respect to $p(x_t \mid y_{1:t-1}, \theta)$ we recover the normal Monte Carlo estimator of that integral. For instance, for the integral in \eqref{eq:dlikfactor} we get:
\begin{align}
  \label{eq:PF:integralApproximation}
  \int_\mathsf{X} p(y_t\mid x_t, \theta) p(x_t \mid
  y_{1:t-1},\theta) \myd x_t
  \approx \int_\mathsf{X} p(y_t\mid x_t, \theta) \left[ \frac{1}{N} \sum_{\ii=1}^N \delta_{x_t^\ii}(x_t) \right]\myd x_t
  = \frac{1}{N} \sum_{\ii=1}^N p(y_t\mid x_t^\ii, \theta).
\end{align}
Let us now consider how the samples---or particles as they are often called---$\{x_t^\ii\}_{\ii=1}^N$ can be generated sequentially.
When $t=1$ we have, by using the convention $y_{1:0} = \emptyset$, that $p(x_1 \mid y_{1:0}, \theta) = p(x_1 \mid \theta)$ and we can thus sample directly from the initial distribution $x_0^\ii \sim \prior(x_0 \mid \theta)$ and simulate the system dynamics to obtain $x_1^\ii \sim p(x_1 \mid x_0^\ii)$ for $\ii = \range{1}{\Np}$.
For any consecutive time step it holds that each target distribution $p(x_{t+1} \mid y_{1:t}, \theta)$ can be constructed from the previous, $p(x_t \mid y_{1:t-1}, \theta)$, in the following way. First, by Bayes' theorem it follows that
\begin{align}
	\label{eq:PF:mu}
	p(x_t \mid y_{1:t}, \theta) \propto p(y_t \mid x_t, \theta) p(x_t \mid y_{1:t-1}, \theta).
\end{align}
Next, by incorporating the state at time $t+1$ (according to the dynamical model) and marginalising over $x_t$ we have
\begin{align}
	\label{eq:PF:tu}
	p(x_{t+1} \mid y_{1:t}, \theta) = \int p(x_{t+1}, x_t \mid
  y_{1:t}, \theta) \myd x_t =
  \int p(x_{t+1} \mid x_t, \theta) p(x_t \mid y_{1:t}, \theta) \myd x_t.
\end{align}
The two equations above are often referred to as the measurement update and time update, respectively, as the former takes the $t^{\text{th}}$ measurement $y_t$ into account and the latter propagates the distribution forward in time according to the system dynamics (cf.\ the two steps of the Kalman filter).

The particle filter makes use of Equations \eqref{eq:PF:mu} and \eqref{eq:PF:tu} to sample approximately from $p(x_{t+1} \mid y_{1:t}, \theta)$ based on the existing samples from $p(x_t \mid y_{1:t-1}, \theta)$. First, we plug the empirical approximation \eqref{eq:PF:EmpiricalApprox} into \eqref{eq:PF:mu}. This gives rise to an empirical approximation of the distribution $p(x_t \mid y_{1:t}, \theta)$, but where each particle is assigned a \emph{weight} $\uw_t^\ii$ according to the multiplicative factor $\uw_t^\ii = p(y_t \mid x_t^\ii, \theta)$. The unknown normalising constant in \eqref{eq:PF:mu} (hidden in the proportionality sign) is not needed since we can simply normalise the weights of the empirical distribution, which becomes
\begin{align}
  \label{eq:PF:EmpiricalAppriximationFilter}
  \widehat{p}(x_t \mid y_{1:t},\theta) = \sum_{\ii=1}^N \frac{\uw_t^\ii}{\sum_{j=1}^N \uw_t^j}\delta_{x^\ii_t}(x_t)  .
\end{align}

Next, we note that by \eqref{eq:PF:tu} it is \emph{conceptually} possible to sample from $p(x_{t+1} \mid y_{1:t}, \theta)$ by first sampling from $p(x_t \mid y_{1:t}, \theta)$ and then simulating the system dynamics forward by sampling from the transition density $p(x_{t+1} \mid x_t, \theta)$.  To make this practical, the particle filter plugs the empirical approximation~\eqref{eq:PF:EmpiricalAppriximationFilter} into~\eqref{eq:PF:tu}, resulting in a \emph{weighted} empirical approximation. Thus, we simulate $N$ particles from the empirical distribution \eqref{eq:PF:EmpiricalAppriximationFilter}. This simply amounts to sampling with replacement from among the existing particles $\{x_t^\ii\}_{\ii=1}^N$, with probabilities given by the normalized weights. Let the resulting particles be denoted by $\{\bar x_t^\ii \}_{\ii=1}^N$. Then we propagate these particles forward in time by sampling $x_{t+1}^\ii \sim p(x_{t+1}^\ii \mid \bar x_t^\ii, \theta)$ for $\ii = \range{1}{N}$. This completes one time step of the particle filter, and the procedure described above can now be repeated for time $t+2$, $t+3$, etc.

The procedure of sampling from the empirical distribution \eqref{eq:PF:EmpiricalAppriximationFilter} is commonly referred to as
\emph{resampling}. This is a key ingredient of the particle filter which, intuitively, exchanges the weight of each particle for a number of copies among the $N$ chosen. Particles with relatively small weights tend to be discarded, while particles with relatively large weights tend to be replicated several times.

We summarize this by Algorithm~\ref{alg:bPF}, the \emph{bootstrap particle filter}, which was introduced independently by \cite{StewartM:1992,Gordon:1993,Kitagawa:1993}.

\begin{algorithm}[ht]
\caption{\textsf{Bootstrap particle filter (all operations are for $\ii=\range{1}{\Np}$)}}
\small
\begin{algorithmic}[1]
  \STATE \textbf{Initialization: }
  \STATE \hspace{4mm} Sample $x_0^{\ii} \sim p(x_0\mid \theta)$ and propagate ${x}^{\ii}_{1} \sim p(x_{1}\mid x_{0}^\ii, \theta)$.
  \STATE \hspace{4mm} Compute $\uw_1^{\ii} = p(y_1\mid {x}_1^\ii, \theta)$.
  \STATE \textbf{for} $t=2$ \textbf{to} $T$ \textbf{do}
  \STATE \hspace{4mm} \textbf{Resampling:} Sample $a_t^{\ii}$ with $\myProb{a_t^{\ii} = j} \propto \uw_{t-1}^{j}$ and set $\bar x_{t-1}^\ii = x_{t-1}^{a_t^{\ii}}$.
 \STATE \hspace{4mm} \textbf{Propagation:} Sample ${x}^{\ii}_{t} \sim
 p(x_{t}\mid \bar x_{t-1}^{\ii}, \theta)$.
  \STATE \hspace{4mm} \textbf{Weighting:} Compute $\uw^\ii_{t} = p(y_{t}\mid {x}_{t}^\ii, \theta)$.
  \STATE \textbf{end}
  \end{algorithmic}
  \label{alg:bPF}
\end{algorithm}

From each empirical approximation $\widehat{p}(x_{t} \mid y_{1:t-1},\theta)$, it is straightforward to perform Monte Carlo integration of (\ref{eq:dlikfactor}) by substituting the empirical approximation $\widehat{p}(x_{t} \mid y_{1:t-1},\theta)$ in place of each $p(x_{t} \mid y_{1:t-1},\theta)$ as in \eqref{eq:PF:integralApproximation}. This results in an estimate of the likelihood given by
\begin{align}
  \label{eq:PF:llest}
  \zhat := \prod_{t=1}^T \left[ \frac{1}{N} \sum_{\ii=1}^N p(y_t\mid x_t^\ii, \theta) \right]
  = \prod_{t=1}^T \left[ \frac{1}{N} \sum_{\ii=1}^N \uw_t^\ii \right].
\end{align}
(Note that the weights $\uw_t^\ii$ appearing in the above expression are
not normalised.) This estimator is obviously non-negative, since each
term $\uw_t^\ii$ is non-negative. What is less obvious to see is that
the estimator also satisfies the requirement of being unbiased, which
means that it can be used within Algorithm~\ref{alg:pseudoMargMP} to
obtain a valid pseudo-marginal Metropolis--Hastings method. For a proof
of this claim we refer the interested reader to \cite{Delmoral:2004,PittSGK:2012}.

\begin{figure}[h]
	\centering
		\setlength{\figureheight}{.2\linewidth}
		\setlength{\figurewidth}{.8\linewidth}
		\pgfplotsset{
			axis on top,
			label style={font=\scriptsize},
			legend style={inner xsep=1pt,inner ysep=0.5pt,nodes={inner sep=1pt,text depth=0.1em},font=\scriptsize},
			tick label style={font=\scriptsize},
		}
		\centering
		\footnotesize
		\input{figures/likelihoodest.tex}
	\caption{Estimation of the likelihood $p(y_{1:T}\mid \theta)$ in the spring-damper example (Section~\ref{sec:NumIll}) for $\theta$ set to the true value. The histograms describe how $10\,000$ independent samples obtained by vanilla Monte Carlo integration (orange, as described in Section~\ref{sec:MCintfp}) and the particle filter (blue, as described in Section~\ref{sec:PF}) are distributed, both using $N=256$. Both approaches provide unbiased estimates (the means, dotted lines, are indeed essentially the same), but the particle filter has significantly smaller variance and less heavy tails than the vanilla Monte Carlo integration (the rightmost orange bin contains all samples $\geq 3$; in fact the biggest orange sample obtained was as large as $100$.) All estimates have been rescaled by a constant factor for clarity of presentation.}
	\label{fig:likelihoodest}
\end{figure}
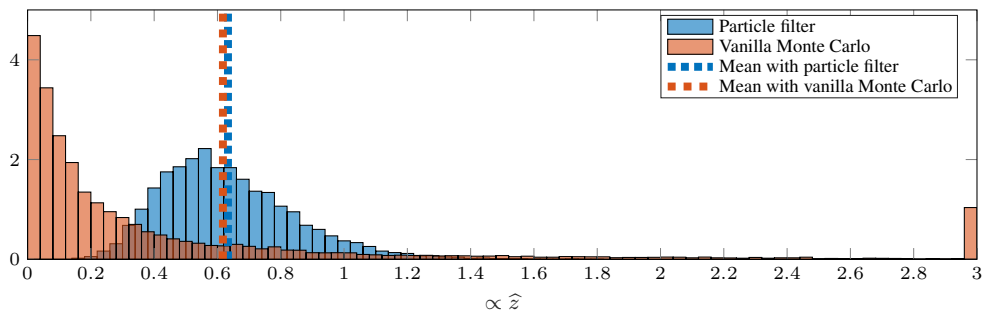

We summarize this section by Figure~\ref{fig:likelihoodest}, where we have implemented the approach proposed in the beginning of Section~\ref{sec:MCintfp} (`vanilla Monte Carlo'), as well as the bootstrap particle filter, to illustrate the different properties of the likelihood estimate $\zhat$ for both approaches. As can be clearly seen, the variance of $\zhat$ obtained from the particle filter is much smaller than the variance obtained by the vanilla Monte Carlo approach. Nevertheless, both approaches provide an unbiased estimate. It should be noted that in more challenging situations the skewness of the distribution of $\zhat$ can be much more extreme (both for the vanilla Monte Carlo approach and for the particle filter if using to few particles) in the sense that $\Prb{\zhat < p(y_{1:T}\mid \theta)} \approx 1$ despite the fact that the estimate is unbiased.

%% file: figures/likelihoodest.tex
%
%
\definecolor{mycolor1}{rgb}{0.00000,0.44700,0.74100}%
\definecolor{mycolor2}{rgb}{0.85000,0.32500,0.09800}%
\begin{tikzpicture}

\begin{axis}[%
width=0.951\figurewidth,
height=\figureheight,
at={(0\figurewidth,0\figureheight)},
scale only axis,
xmin=0,
xmax=3,
xlabel style={font=\color{white!15!black}},
xlabel={$\propto \widehat{z}$},
ymin=0,
ymax=5,
axis background/.style={fill=white},
legend style={legend cell align=left, align=left, draw=white!15!black}
]
\addplot[ybar interval, fill=mycolor1, fill opacity=0.6, draw=black, area legend] table[row sep=crcr] {%
x	y\\
0.02	0\\
0.06	0\\
0.1	0\\
0.14	0.02250450090018\\
0.18	0.0550110022004401\\
0.22	0.16503300660132\\
0.26	0.312562512502501\\
0.3	0.68013602720544\\
0.34	1.00520104020804\\
0.38	1.43028605721144\\
0.42	1.75285057011402\\
0.46	1.85537107421484\\
0.5	2.01790358071614\\
0.54	2.22044408881777\\
0.58	1.8378675735147\\
0.62	1.8378675735147\\
0.66	1.60532106421284\\
0.7	1.36527305461092\\
0.74	1.34026805361072\\
0.78	1.06521304260852\\
0.82	0.952690538107618\\
0.86	0.682636527305462\\
0.9	0.600120024004802\\
0.94	0.47009401880376\\
0.98	0.370074014802961\\
1.02	0.335067013402678\\
1.06	0.257551510302062\\
1.1	0.16503300660132\\
1.14	0.13502700540108\\
1.18	0.117523504700941\\
1.22	0.0675135027005397\\
1.26	0.0725145029005805\\
1.3	0.04750950190038\\
1.34	0.04250850170034\\
1.38	0.02250450090018\\
1.42	0.02750550110022\\
1.46	0.01500300060012\\
1.5	0.0125025005001001\\
1.54	0.01500300060012\\
1.58	0.00500100020004\\
1.62	0.00250050010002002\\
1.66	0.00250050010002\\
1.7	0.00750150030006001\\
1.74	0.00250050010002002\\
1.78	0.00250050010001999\\
1.82	0\\
1.86	0\\
1.9	0\\
1.94	0\\
1.98	0\\
2.02	0\\
2.06	0\\
2.1	0\\
2.14	0\\
2.18	0\\
2.22	0\\
2.26	0\\
2.3	0\\
2.34	0\\
2.38	0\\
2.42	0\\
2.46	0\\
2.5	0\\
2.54	0\\
2.58	0\\
2.62	0\\
2.66	0\\
2.7	0\\
2.74	0\\
2.78	0\\
2.82	0\\
2.86	0\\
2.9	0\\
2.94	0\\
2.98	0\\
3.02	0\\
};
\addlegendentry{Particle filter}

\addplot[ybar interval, fill=mycolor2, fill opacity=0.6, draw=black, area legend] table[row sep=crcr] {%
x	y\\
0	4.48589717943589\\
0.04	3.43818763752751\\
0.08	2.47799559911982\\
0.12	1.94038807761552\\
0.16	1.34776955391078\\
0.2	1.13272654530906\\
0.24	0.955191038207643\\
0.28	0.837667533506701\\
0.32	0.7001400280056\\
0.36	0.552610522104421\\
0.4	0.487597519503901\\
0.44	0.41008201640328\\
0.48	0.360072014402881\\
0.52	0.322564512902581\\
0.56	0.27755551110222\\
0.6	0.27005401080216\\
0.64	0.3000600120024\\
0.68	0.26005201040208\\
0.72	0.210042008401681\\
0.76	0.250050010002\\
0.8	0.18503700740148\\
0.84	0.182536507301461\\
0.88	0.13252650530106\\
0.92	0.13002600520104\\
0.96	0.13252650530106\\
1	0.13002600520104\\
1.04	0.0975195039007801\\
1.08	0.0900180036007206\\
1.12	0.0775155031006196\\
1.16	0.0825165033006605\\
1.2	0.0775155031006196\\
1.24	0.0800160032006405\\
1.28	0.06751350270054\\
1.32	0.0725145029005801\\
1.36	0.0700140028005601\\
1.4	0.0625125025005\\
1.44	0.0650130026005204\\
1.48	0.0750150030006005\\
1.52	0.05251050210042\\
1.56	0.0625125025004997\\
1.6	0.0425085017003405\\
1.64	0.0425085017003398\\
1.68	0.05501100220044\\
1.72	0.0525105021004203\\
1.76	0.0500100020004\\
1.8	0.0500100020004\\
1.84	0.0325065013002602\\
1.88	0.0375075015002998\\
1.92	0.0375075015003\\
1.96	0.0400080016003203\\
2	0.0450090018003603\\
2.04	0.0425085017003396\\
2.08	0.0300060012002404\\
2.12	0.04500900180036\\
2.16	0.03000600120024\\
2.2	0.02750550110022\\
2.24	0.02250450090018\\
2.28	0.0375075015003\\
2.32	0.02250450090018\\
2.36	0.03000600120024\\
2.4	0.03000600120024\\
2.44	0.04250850170034\\
2.48	0.01000200040008\\
2.52	0.0175035007001402\\
2.56	0.0150030006001198\\
2.6	0.01500300060012\\
2.64	0.0225045009001803\\
2.68	0.02000400080016\\
2.72	0.01750350070014\\
2.76	0.01500300060012\\
2.8	0.00500100020004\\
2.84	0.01500300060012\\
2.88	0.00750150030006001\\
2.92	0.01500300060012\\
2.96	1.0377075415083\\
3	1.0377075415083\\
};
\addlegendentry{Vanilla Monte Carlo}

\addplot [color=mycolor1, dotted, line width=3.0pt]
  table[row sep=crcr]{%
0.633226214999029	0\\
0.633226214999029	5\\
};
\addlegendentry{Mean with particle filter}

\addplot [color=mycolor2, dashed, line width=3.0pt]
  table[row sep=crcr]{%
0.617221165067179	0\\
0.617221165067179	5\\
};
\addlegendentry{Mean with vanilla Monte Carlo}

\end{axis}
\end{tikzpicture}%

%% file: pmh.tex
\section{Using the particle filter estimate within pseudo-marginal MH}
\label{sec:PMH}
The non-negative and unbiased likelihood estimator described in
Section~\ref{sec:PF} can now be used inside the pseudo-marginal
Metropolis--Hastings algorithm from Section~\ref{sec:pmMCMC}. The end
result is a standard Metropolis--Hastings algorithm operating on the
non-standard joint space of the model parameters $\theta$ and the
auxiliary variable $\zhat$. The target distribution for our new algorithm
is given by $\target(\theta, z\mid y_{1:\T})$, which was defined
in~\eqref{eq:PMMH:ext-target-willwork}. The result is referred to as
\emph{particle Metropolis--Hastings (PMH)}---summarized in
Algorithm~\ref{alg:PMH}. It was introduced by Andrieu et al. in
$2010$~\cite{AndrieuDH:2010} (under the name particle \emph{marginal} Metropolis--Hastings).
\begin{algorithm}[ht]
\caption{\textsf{Particle Metropolis--Hastings (PMH)}}
\small
\begin{algorithmic}[1]
  \STATE \textbf{Initialisation ($m=0$):} Set $\theta[0]$ arbitrarily
  and run Algorithm~\ref{alg:bPF} to obtain $\zhat[0]$.
  \STATE \textbf{for} $m = 1$ \textbf{to} $M$ \textbf{do}
  \STATE \hspace{4mm} Sample $\theta^{\prime} \sim q(\theta \mid \theta[m-1])$.
  \STATE \hspace{4mm} Sample $\zhat^{\prime} \sim \psi(z \mid \theta^{\prime},
  y_{1:\T})$ by running Algorithm~\ref{alg:bPF} once and
  compute~\eqref{eq:PF:llest}.
  \STATE \hspace{4mm} Compute the acceptance probability
  \begin{align}
  \label{eq:alg:PMH:ap}
    \alpha_m = \min\left(1,\frac{\zhat^{\prime}\prior(\theta^{\prime})}{\zhat[m-1]\prior(\theta[m-1])}
    \frac{q(\theta[m-1]\mid \theta^\prime)}{q(\theta^\prime\mid \theta[m-1])}\right),
  \end{align}
  \STATE \hspace{4mm} Sample $\omega_m$ uniformly over $[0, 1]$.
  \STATE \hspace{4mm} \textbf{if} $\omega_m < \alpha_m$ \textbf{then}
  \STATE \hspace{4mm} \hspace{4mm} Set $\{\theta[m], \zhat[m]\} \leftarrow \{\theta^{\prime}, \zhat^{\prime}\}$ (accept the candidate samples).
  \STATE \hspace{4mm} \textbf{else}
  \STATE \hspace{4mm} \hspace{4mm} Set $\{\theta[m], \zhat[m]\} \leftarrow \{\theta[m-1], \zhat[m-1]\}$ (reject the candidate samples).
  \STATE \hspace{4mm} \textbf{end}
  \STATE \textbf{end}
\end{algorithmic}
\label{alg:PMH}
\end{algorithm}

The PMH algorithm will thus produce a Markov chain
$\{\theta[m], \zhat[m]\}_{m=0}^{M}$ on the joint space of the parameters
$\theta$ and the auxiliary variable $\zhat$.  Since the posterior
distribution $p(\theta\mid y_{1:\T})$ is obtained as a marginal of
$\target(\theta, z\mid y_{1:\T})$ we can---by construction---obtain
samples from $p(\theta\mid y_{1:\T})$ by extracting the sub-chain
$\{\theta[m]\}_{m=0}^{M}$ from $\{\theta[m], \zhat[m]\}_{m=0}^{M}$. Hence,
the samples of the auxiliary variable are simply ignored.
It is worth noting that the samples of the states $x_{0:\T}$ produced
by the particle filters that are executed as parts of
Algorithm~\ref{alg:PMH} provide a competitive solution to the problem
of estimating the joint smoothing distribution $p(x_{0:\T}\mid
y_{1:\T})$.

As noted in Section~\ref{sec:ParameterInference}, the variance of the estimate $\zhat$ will affect the convergence speed of the algorithm. Specifically, if the variance is very large, then the method tends to get ``stuck'' for many consecutive iterations, not accepting any proposed moves. The reason for this is that large variance in the estimate is often related to a large skewness as well, in the sense that with high probability $\zhat < p(y_{1:T}\mid \theta)$. However, since the estimate is unbiased, this implies that $\zhat$ sometimes (but rarely) must take on values $\zhat \gg p(y_{1:T}\mid \theta)$. This is illustrated by Figure~\ref{fig:likelihoodest} where the vanilla Monte Carlo estimate is heavily skewed.  
This is a problem when $\zhat$ is used in Algorithm~\ref{alg:PMH} since if most of the probability mass is concentrated on $\zhat \ll p(y_{1:T}\mid \theta)$, the acceptance probability $\alpha_m$ will tend to be small and the sampler can get stuck for many iterations.
For the method to work satisfactorily we therefore need to ensure that the variance in $\zhat$ is not overly large, which in turn implies that we need to use a sufficiently large number $N$ of particles in the underlying particle filter. What ``sufficiently'' means is problem dependent, but as a rule-of-thumb $N$ should scale linearly with $T$.

%
A possible improvement of Algorithm~\ref{alg:PMH} is offered by noting that
the particle filter output can also be used to compute estimates of the
likelihood gradient and Hessian. These estimates can be used to
construct better proposal distributions $q$ for $\theta$, with the potential to explore
the parameter space similarly to the way in which Newton's
optimization algorithm is exploring the parameter space. This can lead to significant improvements compared to the standard random walk
proposal that is commonly employed, see \cite{DahlinLS:2015}. The
classic variance reduction technique of positively correlating two
estimators has been applied also to the particle Metropolis--Hastings
algorithm by introducing positive correlations between subsequent
likelihood estimators~\cite{DeligiannidisDP:2015}.

%% file: varRedPF.tex
\section{Variance reduction methods for the particle filter}
\label{sec:VarRedPF}
As illustrated by Figure~\ref{fig:likelihoodest}, the bootstrap particle filter results in a substantial improvement in the likelihood estimate over the vanilla Monte Carlo approach. Since both approaches result in unbiased estimates, this improvement essentially stems from a reduction in variance.
Various improvements on the bootstrap particle filter can similarly be used to further reduce the variance of the estimate of the likelihood (while maintaining unbiasedness). Indeed, many such improvements have been proposed in the literature over the past quarter century. For the interested reader we review a few key ideas below. Additional improvements and details can be found in e.g.\ \cite{DoucetJ:2011,CappeGM:2007,DoucetGA:2000}.

\subsection{Low-variance resampling}
Firstly, we note that the resampling step of the particle filter---while being of key importance to its stability---is a random procedure. As such it inevitably introduces some additional variance in the estimate of the likelihood. This additional variance can be reduced by using standard variance reduction techniques, e.g.\ stratification, when sampling from the empirical distribution \eqref{eq:PF:EmpiricalAppriximationFilter}. This still results in a valid particle filter, as long as the resampling method used is itself unbiased. More precisely, when sampling from the empirical distribution \eqref{eq:PF:EmpiricalAppriximationFilter} we require that the expected number of copies of each particle is proportional to its weight,
\begin{align}
  \Exp{\frac{1}{N}\sum_{\ii=1}^N I(\bar x_{t}^\ii = x_{t}^i)} &= \frac{w_{t}^i}{\sum_{\ii=1}^N w_t^\ii}, & i&=\range{1}{N},
\end{align}
where $I(\cdot)$ is an indicator function. Several low-variance resampling methods satisfying this unbiasedness condition are reviewed in \cite{DoucCM:2005,HolSG:2006}.

Low-variance resampling can (and should) always be used when implementing the particle filter.
Other variance reduction techniques, however, are more model-specific and can be used only for certain classes of state-space models. We describe two such possible improvements below.

\subsection{Conditioning on the current measurement}
The first such improvement is a technique which aims to take the current measurement $y_t$ into account when simulating the particles $\{x_t^\ii\}_{\ii=1}^{\Np}$ at time $t$. The intuition behind this is that the measurement $y_t$ often contains valuable information about the state of the system at time $t$, which can help to simulate these particles in a ``good'' region of the state space. To this end we assume that the distribution for the system dynamics \emph{conditional} on the current observation---i.e.
\begin{align}
	\label{eq:APF:optimal_q}
	p(x_t \mid x_{t-1}, y_t, \theta) = \frac{p(y_t \mid x_t, \theta) p(x_t \mid x_{t-1}, \theta)}{p(y_t \mid x_{t-1}, \theta)},
\end{align}
---is available in closed form.
This is not always the case, but for some highly relevant state-space models it is indeed available. One important example is when the state dynamics $p(x_t \mid x_{t-1}, \theta)$ are Gaussian (but with a possibly nonlinear dependence on $x_{t-1}$) and the measurement equation is linear and Gaussian, $p(y_t \mid x_t, \theta) = \mathcal{N}(y_t \mid Cx_t, R)$, for the (non state dependent) matrices $C$ and $R$. We shall further assume that the normalization factor $p(y_t \mid x_{t-1}, \theta)$ in \eqref{eq:APF:optimal_q} can be evaluated point-wise, but this typically follows from the aforementioned assumption.

The effect of simulating particles $x_t^\ii$ conditionally on $y_t$ is that we obtain an (unweighted) empirical approximation of the \emph{filtering distribution} instead of the one-step predictive distribution as in~\eqref{eq:PF:integralApproximation}. That is, we sequentially obtain the empirical approximations
\begin{align}
  \label{eq:APF:EmpiricalApprox}
  \widehat{p}(x_t \mid y_{1:t},\theta) = \frac{1}{N}\sum_{\ii=1}^N \delta_{x^\ii_t}(x_t) ,
\end{align}
for $t=\range{0}{T}$. Note the conditioning on $y_t$ on the left-hand side in~\eqref{eq:APF:EmpiricalApprox}.

Our main object of interest---the likelihood---can be expressed in terms of these filtering distributions using a similar factorisation as in \eqref{eq:dlikfactor}, but where we choose to marginalise over $x_{t-1}$ instead of $x_t$,
\begin{align}
p(y_{1:T} \mid \theta) = \prod_{t=1}^{T} \int_\mathsf{X} p(y_t\mid x_{t-1}, \theta) p(x_{t-1} \mid
y_{1:t-1},\theta) \myd x_{t-1}.
\label{eq:APF:dlikfactor}
\end{align}

To see how we can sequentially obtain the empirical distributions \eqref{eq:APF:EmpiricalApprox} we start by combining the measurement and time update equations, \eqref{eq:PF:mu} and \eqref{eq:PF:tu}, with the expression \eqref{eq:APF:optimal_q},
\begin{align}
  \notag
  p(x_{t} \mid y_{1:t}, \theta) &\propto
                                  \int p(y_t \mid x_t, \theta) p(x_{t} \mid x_{t-1}, \theta) p(x_{t-1} \mid y_{1:t-1}, \theta) \myd x_{t-1} \\
                                &=
	\int p(x_{t} \mid x_{t-1}, y_t, \theta) p(y_t \mid x_{t-1}, \theta) p(x_{t-1} \mid y_{1:t-1}, \theta) \myd x_{t-1}.
\end{align}
Proceeding in a similar way as for the bootstrap particle filter presented above, we plug in the ``current'' empirical approximation of the filtering distribution (i.e., from time $t-1$) into the above integral.
In doing so, the factor $p(y_t \mid x_{t-1}, \theta)$ enters as a weight on the particles at time $t-1$. Thus, to sample a new set of particles $\{x_t^\ii\}_{\ii=1}^N$ approximately from the filtering distribution at time $t$, we \emph{(i)} resample the particles at time $t-1$ with probabilities proportional to $\nu_{t}^\ii := p(y_t \mid x_{t-1}^\ii, \theta)$, yielding $\{\bar x_{t-1}^\ii\}_{\ii=1}^N$ and \emph{(ii)} propagate these particles to time~$t$ by sampling $x_t^\ii \sim p(x_t \mid \bar x_{t-1}^\ii, y_t , \theta)$ for $\ii = \range{1}{N}$.

Finally, by \eqref{eq:APF:dlikfactor}, the estimate of the likelihood is given by
\begin{align}
	\label{eq:APF:llest}
	z := \prod_{t=1}^T \left[ \frac{1}{N} \sum_{\ii=1}^N \nu_t^\ii \right].
\end{align}
As for the bootstrap particle filter, it holds that this estimator is non-negative and unbiased.
However, it is often the case in practice that \eqref{eq:APF:llest} has much lower variance than its bootstrap particle filter counterpart~\eqref{eq:PF:llest}.

The particle filter described above is often referred to as the \emph{fully adapted auxiliary particle filter}. The term ``adapted'' here refers to the fact that we adapt the sampling distributions, both in the resampling step and when simulating new particles, to the information provided by the current measurement $y_t$. ``Fully'' refers to the fact that we do so by using the exact conditional distribution in \eqref{eq:APF:optimal_q}.
As mentioned above, however, this conditional distribution is only available in closed form for a restricted class of models. More generally it is possible to use an approximation $q(x_t \mid x_{t-1}, y_t, \theta) \approx p(x_t \mid x_{t-1}, y_t, \theta)$ for simulating new particles, and similarly an approximation $q(y_t \mid x_{t-1}, \theta) \approx p(y_t \mid x_{t-1}, \theta)$ when computing the resampling weights. It turns out that it is still possible to obtain a valid particle filter implementation,  yielding unbiased estimates of the likelihood, where the approximations $q$ are used as \emph{proposal distributions} within an importance sampling framework. We refer to \cite{PittS:1999,DoucetJ:2011} for details. A complete textbook-style introduction along the lines of this section is available in \citep{SchonL:2017}.

\subsection{Rao-Blackwellization}
One important (and rather common) class of state-space models for which another variance reduction technique is possible are the so-called \emph{conditionally linear Gaussian (CLG)} state-space models. These models are characterised by having a substructure that can be identified as being linear and Gaussian. Therefore, this substructure is analytically tractable using Kalman filtering techniques, which can be leveraged when running a particle filter for the whole model. This is done in a way which resembles the method of \emph{Rao-Blackwellization} of statistical estimators, and the resulting particle filters are therefore commonly referred to as Rao-Blackwellized particle filters (RBPFs).

To be more specific, let the state variable $x_t$ be partitioned into two components $x_t = (x_t^\text{n}, x_t^\text{l})$ where we use the superscripts n and l for nonlinear and linear, respectively. Then, a CLG model can be defined as a state-space model where the \emph{conditional stochastic process} $\{(x_t^\text{l}, y_t) \mid x_{0:t}^\text{n}\}_{t\geq 0}$ follows a (time-inhomogeneous) linear Gaussian state-space model. We can thus identify whether or not a specific model under study is CLG by ``pretending'' that the nonlinear state component $\{x_t^\text{n}\}_{t\geq 0}$ is observed. If the model then can be viewed as a linear Gaussian state-space model, then the original model is CLG.

To give an example, consider the mixed linear/nonlinear model
\begin{subequations}
  \label{eq:PBPF:MLN}
  \begin{align}
    \label{eq:PBPF:MLN:a}
    x_{t+1}^\text{n} &= f^\text{n}(x_t^\text{n}) + A^\text{n}(x_t^\text{n})x_t^\text{l} + v_t^\text{n}, \\
    x_{t+1}^\text{l} &= f^\text{l}(x_t^\text{n}) + A^\text{l}(x_t^\text{n})x_t^\text{l} + v_t^\text{l}, \\
    y_t &= g(x_t^\text{n}) + C(x_t^\text{n})x_t^\text{l} + e_t,
  \end{align}
  \end{subequations}
where $x_t^\text{n} \in \reals^{d_\text{n}}$, $x_t^\text{l} \in \reals^{d_\text{l}}$, and $y_t \in \reals^{d_y}$. The model is specified in terms of the (possibly nonlinear) functions
$f^\text{n}: \reals^{d_\text{n}}\mapsto \reals^{d_\text{n}}$,
$A^\text{n}: \reals^{d_\text{n}}\mapsto \reals^{d_\text{n}\times d_\text{l}}$,
$f^\text{l}: \reals^{d_\text{n}}\mapsto \reals^{d_\text{l}}$,
$A^\text{l}: \reals^{d_\text{n}}\mapsto \reals^{d_\text{l}\times d_\text{l}}$,
$g: \reals^{d_\text{n}}\mapsto \reals^{d_y}$, and
$C: \reals^{d_\text{n}}\mapsto \reals^{d_y\times d_\text{l}}$
(where we have suppressed the dependence on the model parameter $\theta$ for brevity). Furthermore, the process noises $v_t^\text{n}$ and $v_t^\text{l}$, as well as the measurement noise $e_t$ are all assumed to be Gaussian. To see that this model is indeed CLG, pretend that $\{x_t^\text{n}\}_{t\geq0}$ is observed. Then, all the functions just listed (which specify the model) are known constants and the model is reduced to a linear Gaussian state-space model. Note that the dynamical equation \eqref{eq:PBPF:MLN:a} is viewed as a measurement equation in this ``pretend'' linear Gaussian state-space model.

The RBPF exploits the structure of a CLG model by simulating particles representing the nonlinear state $x_t^\text{n}$, while simultaneously tracking the linear state $x_t^\text{l}$ using a separate Kalman filter for each particle.\footnote{To be more precise, the RBPF maintains particles representing the entire history of the nonlinear state $x_{0:t}^\text{n}$ and conditional on these histories the model is linear subject to Gaussian noise.} We do not go in to the details on the RBPF implementation here, but instead refer the interested reader to \cite{ChenL:2000,DoucetGA:2000}.
Note that CLG models can come in many different forms (hence the rather abstract definition above). However, the variance reduction offered by Rao-Blackwellization can in many cases be substantial, so it is well worth the effort to investigate whether or not a given model under study can be viewed as CLG---thus opening up for using an RBPF---before running a bootstrap particle filter on this model.
Automatic Rao-Blackwellization is a recent topic of research in probabilistic programming \citep{MurrayLKBS:2017}.

%% file: numIll.tex
\section{Numerical illustrations: Learning a nonlinear spring-damper system}
\label{sec:NumIll}
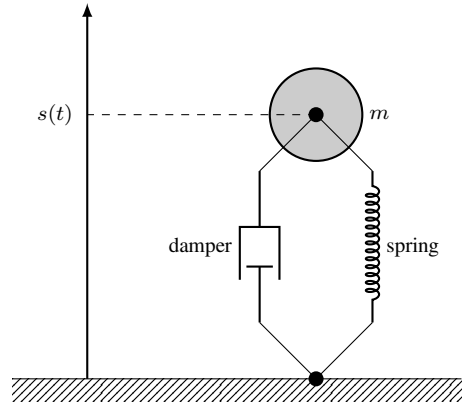
\begin{figure}[t]
	\centering
	\setlength{\figureheight}{.2\linewidth}
	\setlength{\figurewidth}{.8\linewidth}
	\pgfplotsset{
		axis on top,
		label style={font=\scriptsize},
		legend style={inner xsep=1pt,inner ysep=0.5pt,nodes={inner sep=1pt,text depth=0.1em},font=\scriptsize},
		tick label style={font=\scriptsize},
	}
	\centering
	\footnotesize
	\input{figures/springdamper.tex}
	\caption{The spring-damper system.}
	\label{fig:system}
\end{figure}

\begin{figure}[t]
	\centering
		\setlength{\figureheight}{.2\linewidth}
		\setlength{\figurewidth}{.8\linewidth}
		\pgfplotsset{
			axis on top,
			label style={font=\scriptsize},
			legend style={inner xsep=1pt,inner ysep=0.5pt,nodes={inner sep=1pt,text depth=0.1em},font=\scriptsize},
			tick label style={font=\scriptsize},
			/pgfplots/area legend/.style={
				legend image code/.code={
					\fill[##1]             (.25cm,-0.1cm) rectangle (0.35cm,0.1cm);
					\fill[##1,opacity=0.3] (.20cm,-0.1cm) rectangle (0.40cm,0.1cm);
					\fill[##1,opacity=0.3] (.15cm,-0.1cm) rectangle (0.45cm,0.1cm);
					\fill[##1,opacity=0.3] (.10cm,-0.1cm) rectangle (0.50cm,0.1cm);
					\fill[##1,opacity=0.3] (.05cm,-0.1cm) rectangle (0.55cm,0.1cm);
					\fill[##1,opacity=0.3] (.00cm,-0.1cm) rectangle (0.60cm,0.1cm);
				}
			}
		}
		\centering
		\footnotesize
		\input{figures/data.tex}
	\caption{Data simulated from the spring-damper system, which will be used to learn the parameter values $\theta$. As clearly can be seen, there is a significant amount of stochastic noise present in the system, and the data record is not very large.}
	\label{fig:data}
\end{figure}
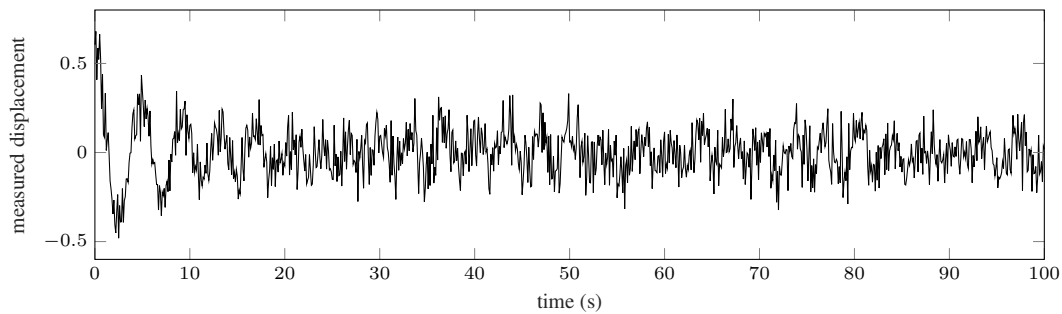

\begin{figure}[t]
	\centering
		\setlength{\figureheight}{.4\linewidth}
		\setlength{\figurewidth}{\linewidth}
		\pgfplotsset{
			axis on top,
			label style={font=\scriptsize},
			legend style={inner xsep=1pt,inner ysep=0.5pt,nodes={inner sep=1pt,text depth=0.1em},font=\scriptsize},
			tick label style={font=\scriptsize},
		}
		\centering
		\footnotesize
		\input{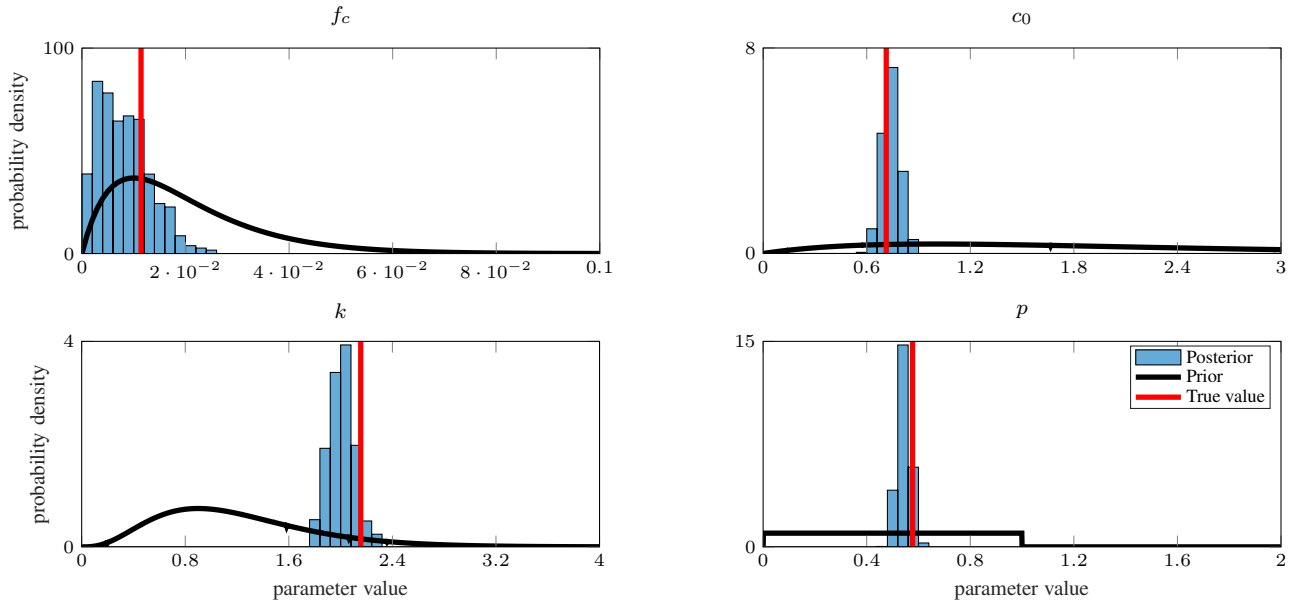}
		\caption{The marginals of the posterior distribution (blue histograms) for the parameters $\theta$, i.e., what can be said about the parameters given measurements $y_{1:T}$ in Figure~\ref{fig:data}, the priors (black curves) and the model~\eqref{eq:ex:model} illustrated using vertical lines at the true parameter values.}
		\label{fig:posterior1}
\end{figure}

In this section, we will walk through a basic but
illustrative example of probabilistic learning in an applied
situation. The code used in the example is available in the appendix
and also via the first author's homepage. For a more challenging
numerical example, we refer to, e.g., \cite{SvenssonLS:2017abc}. The
nonlinear spring-damper system that is used to exemplify the method
is illustrated in Figure~\ref{fig:system} and it is modeled via a
spring force $F_{\text{s}}$ and a damper force $F_{\text{d}}$
according to
\begin{subequations}
	\label{eq:ex:model}
	\begin{align}
    	F_{\text{s}} &= -ks^p,\\
	    F_{\text{d}} &= -f_{\text{c}}\text{sign}(\dot{s})-c_0\dot{s},
	\end{align}
\end{subequations}
where $k$ denotes the spring coefficient and $c_0$ denotes the damper coefficient. Furthermore $\dot{s}$ denotes the derivative of the displacement $s$ with respect to time. The spring will be linear for $p=1$, while $p<1$ results in a nonlinear spring. Furthermore, $f_{\text{c}}=0$ gives a linear damper, and the nonlinear damper $f_{\text{c}}>0$ is motivated in \cite{SDS+:1997}. Via the use of force balance $\sum F = m\ddot{s}$ and a forward-Euler discretization with sampling time $T_{\text{s}}$ we obtain the following discrete-time state-space model
\begin{subequations}
  \begin{align}
    x^1_{t+1} &= x_t^1 + T_{\text{s}} x_t^2,\\
    x^2_{t+1} &= x_t^2 + \tfrac{T_{\text{s}}}{m}(-f_{\text{c}}\text{sign}(x_t^2)-c_0x_t^2-k(x_t^1)^p) + v_t,\\
    y_t &= x_t^1 + e_t,
  \end{align}
\end{subequations}
where $x_t^1$ represents $s(t)$ and $x_t^2$ represents $\dot{s}(t)$, and we have also included measurement noise $e_t$ and process noise $v_t$ (e.g. an external unknown force). From this model, $T=1\,000$ data points were simulated with an initial known mass displacement $s(0) = 0.5$, shown in Figure~\ref{fig:data}. Furthermore, the sampling time was $T_{\text{s}} = 0.1$, the mass $m=8$, and the noise distributions $e_t\sim \mathcal{N}(0,0.1^2)$ and $v_t\sim \mathcal{N}(0,0.01^2)$, which we assume is known, whereas the remaining parameters $\theta = (k, p, f_c, c_0)^{\Transp}$ are assumed unknown. Their true values, as used in the simulation, are as follows: $k^{\star} = 2.16$, $p^{\star} = 0.58$, $f_c^{\star} = 0.01$ and $c_0^{\star} = 0.71$ (see also Figure~\ref{fig:posterior1}).

Let us now specify the priors to be used. The parameters
$f_{\text{c}}, c_0$ and $k$ all have to be non-negative by
construction, implying that any reasonable prior must be non-negative
as well. The Gamma distribution---which we will denote by
$\mathcal{G}$---is a reasonable choice. More
specifically, since $f_{\text{c}}$ is likely to be small we choose
$f_{\text{c}} \sim \mathcal{G}(2,0.01)$, $c_0$ on the other hand can
take on slightly larger values, so its prior is chosen as
$\mathcal{G}(2,1)$. For the spring coefficient $k$ the prior is chosen
as $\mathcal{G}(4,.3)$. Finally, $p$ is by physical laws restricted to the interval
$0 \leq p \leq 1$, which motivates a uniform prior $\mathcal{U}(0,1)$. These
priors for the parameters are shown in Figure~\ref{fig:posterior1}. 

\begin{figure}[t]
	\centering
		\setlength{\figureheight}{.2\linewidth}
		\setlength{\figurewidth}{.8\linewidth}
		\pgfplotsset{
			axis on top,
			label style={font=\scriptsize},
			legend style={inner xsep=1pt,inner ysep=0.5pt,nodes={inner sep=1pt,text depth=0.1em},font=\scriptsize},
			tick label style={font=\scriptsize},
			/pgfplots/area legend/.style={
				legend image code/.code={
					\fill[##1]             (.25cm,-0.1cm) rectangle (0.35cm,0.1cm);
					\fill[##1,opacity=0.3] (.20cm,-0.1cm) rectangle (0.40cm,0.1cm);
					\fill[##1,opacity=0.3] (.15cm,-0.1cm) rectangle (0.45cm,0.1cm);
					\fill[##1,opacity=0.3] (.10cm,-0.1cm) rectangle (0.50cm,0.1cm);
					\fill[##1,opacity=0.3] (.05cm,-0.1cm) rectangle (0.55cm,0.1cm);
					\fill[##1,opacity=0.3] (.00cm,-0.1cm) rectangle (0.60cm,0.1cm);
				}
			}
		}
		\centering
		\footnotesize
		\input{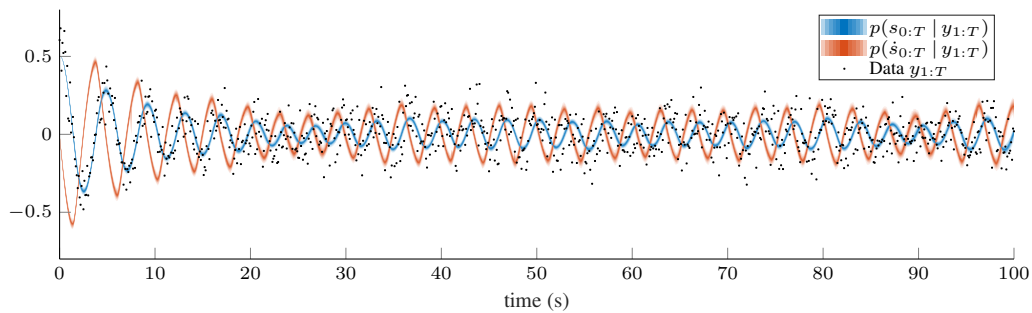}
	\caption{The measurements (dots) alongside with the computed state smoothing distributions, i.e., $p(s_{0:T}\mid y_{1:T})$ (shaded blue) and $p(\dot{s}_{0:T}\mid y_{1:T})$ (shaded orange).}
	\label{fig:states}
\end{figure}

With the model definition $p(\theta, x_{0:\T}, y_{1:\T})$ in place we
can now start to learn~$\theta$. We did this by generating
$10\thinspace 000$ samples from $p(\theta\mid y_{1:T})$ using PMH as
described in Algorithm~\ref{alg:PMH}. The result is shown in
Figure~\ref{fig:posterior1} (together with the true values and the
priors). As the proposal in PMH, we used a random walk based on a
Gaussian distribution with rather small variance. This takes a few
minutes on an ordinary desktop computer. As we have mentioned the PMH
algorithm will also provide samples from the state smoothing
distribution $p(x_{0:T}\mid y_{1:T})$, which we show in
Figure~\ref{fig:states}.

%% file: figures/springdamper.tex


  \begin{tikzpicture}[every node/.style={draw,outer sep=0pt,thick}]
    \tikzstyle{spring}=[thick,decorate,decoration={coil,pre length=0.2cm,post length=0.2cm,segment length=3}]
    \tikzstyle{damper}=[thick,
                        decoration={markings,
                                    mark connection node=dmp,
                                    mark=at position 0.5 with {
                                                                \node (dmp) [thick,inner sep=0pt,transform shape,rotate=-90,minimum width=15pt,minimum height=15pt,draw=none] {};
                                                                \draw [thick] ($(dmp.north east)+(5pt,0)$) -- (dmp.south east) -- (dmp.south west) -- ($(dmp.north west)+(5pt,0)$);
                                                                \draw [thick] ($(dmp.north)+(0,-5pt)$) -- ($(dmp.north)+(0,5pt)$);
                                                              }
                                   },
                        decorate]
    \tikzstyle{ground}=[fill,pattern=north east lines,draw=none,minimum width=0.75cm,minimum height=0.3cm]
    
    \node (M) [circle,label=right:$m$,minimum width=35,fill=black!20]{};
    \fill circle[fill=black,radius=.1];
    \node (ground) [ground,anchor=north,xshift=-30,yshift=-100,minimum width=170] at (M){};
    \draw (ground.north east) -- (ground.north west);
    
    \draw (M.center) -- ++(.75,-.75) coordinate (springstart);
    \draw (M.center) -- ++(-.75,-.75) coordinate (damperstart);
    \fill (M.center|-ground.north) circle[radius=.1];
    \draw (M.center|-ground.north) -- ++ (.75,.75) coordinate (springend);
    \draw (M.center|-ground.north) -- ++ (-.75,.75) coordinate (damperend);
    \draw[spring] (springstart) -- (springend) node[draw=none,midway,label=right:spring]{};
    \begin{scope}[label distance = 5];
    \draw[damper] (damperstart) -- (damperend) node[draw=none,midway,label=left:damper]{};
    \end{scope}
    \draw [-latex,thick] (ground.north west) ++ (1,0) -- ++(0,5) coordinate (arrowtop);
    \draw [dashed] (M.center) -- (arrowtop|-M.center) node[draw=none,label=left:$s(t)$]{};
    
  \end{tikzpicture}

%% file: figures/data.tex
%
%
\begin{tikzpicture}

\begin{axis}[%
width=0.951\figurewidth,
height=\figureheight,
at={(0\figurewidth,0\figureheight)},
scale only axis,
xmin=0,
xmax=100,
xlabel style={font=\color{white!15!black}},
xlabel={time (s)},
ymin=-0.6,
ymax=0.8,
ylabel style={font=\color{white!15!black}},
ylabel={measured displacement},
axis background/.style={fill=white}
]
\addplot [color=black, forget plot]
  table[row sep=crcr]{%
0	0.604875233303903\\
0.1	0.680117483933636\\
0.2	0.408446072128873\\
0.3	0.586709538769557\\
0.4	0.52253101473142\\
0.5	0.66395424365665\\
0.6	0.52959094456363\\
0.7	0.245979912812654\\
0.8	0.438832832611936\\
0.9	0.101432468258714\\
1	0.0859610853635491\\
1.1	0.333064014934465\\
1.2	0.0115570869376518\\
1.3	0.0674936523114132\\
1.4	0.111638729668267\\
1.5	-0.0185186883623662\\
1.6	-0.159548151793011\\
1.7	-0.205203540056103\\
1.8	-0.262569375002154\\
1.9	-0.345062367454069\\
2	-0.267124142992121\\
2.1	-0.404840131849161\\
2.2	-0.451722698154678\\
2.3	-0.331342827580425\\
2.4	-0.236091160266401\\
2.5	-0.480944737110418\\
2.6	-0.297914152414715\\
2.7	-0.389960254694061\\
2.8	-0.388261497098232\\
2.9	-0.202519736473365\\
3	-0.391872870629313\\
3.1	-0.270186109425536\\
3.2	-0.247556394399027\\
3.3	-0.148367184680453\\
3.4	-0.147378933406961\\
3.5	-0.166539568278606\\
3.6	-0.0559465770374565\\
3.7	-0.00370843007809176\\
3.8	-0.00103236012535347\\
3.9	0.135510996060095\\
4	0.217826869247276\\
4.1	0.239312726684035\\
4.2	0.0758591723867104\\
4.3	0.0936451209730466\\
4.4	0.100397819216265\\
4.5	0.329119680673487\\
4.6	0.267289220628544\\
4.7	0.329256693881281\\
4.8	0.0632617709886026\\
4.9	0.435077388761324\\
5	0.345885755516321\\
5.1	0.261392747069917\\
5.2	0.145366726590664\\
5.3	0.294660600589396\\
5.4	0.188264586801037\\
5.5	0.246370115896549\\
5.6	0.199953017698106\\
5.7	0.091443917339988\\
5.8	0.233567861765771\\
5.9	0.0383638469854314\\
6	-0.0650016703537401\\
6.1	-0.039299494196334\\
6.2	-0.0626728809227695\\
6.3	-0.0568577934508397\\
6.4	-0.121253033682481\\
6.5	-0.179796839006905\\
6.6	-0.180299968364358\\
6.7	-0.322817704026371\\
6.8	-0.185785314288406\\
6.9	-0.22227242900679\\
7	-0.354898432237873\\
7.1	-0.223208182761283\\
7.2	-0.241696875020134\\
7.3	-0.201439548595624\\
7.4	-0.308034957545387\\
7.5	-0.184626282252005\\
7.6	-0.161986015053706\\
7.7	-0.219949914792128\\
7.8	-0.041894744250051\\
7.9	-0.215317800874161\\
8	-0.00957651532663401\\
8.1	-0.05271931173631\\
8.2	0.0757736323912553\\
8.3	0.0789852683490601\\
8.4	0.117306224168069\\
8.5	-0.0154288470015478\\
8.6	0.344373749113857\\
8.7	0.0545743764811382\\
8.8	0.10697626899442\\
8.9	0.0671460203287409\\
9	0.241983305403104\\
9.1	0.132323469065027\\
9.2	0.144004088827677\\
9.3	0.240914209257562\\
9.4	0.244736206923387\\
9.5	0.289259890366279\\
9.6	-0.0161435722810859\\
9.7	0.211962036802169\\
9.8	0.120585557499493\\
9.9	0.0792854407449138\\
10	0.151590365500779\\
10.1	0.158474757193411\\
10.2	0.108485669923334\\
10.3	-0.113988958265525\\
10.4	-0.0014133106961287\\
10.5	-0.111748844923711\\
10.6	-0.0771517672585238\\
10.7	-0.0995658273297718\\
10.8	0.0471908256986399\\
10.9	-0.126932119177945\\
11	-0.264945497102425\\
11.1	-0.0527384390716339\\
11.2	-0.156105388539446\\
11.3	-0.187735561569855\\
11.4	-0.138508189048829\\
11.5	-0.148688907444302\\
11.6	-0.0704736292810097\\
11.7	-0.117117392480371\\
11.8	0.0532977798136356\\
11.9	-0.120179815283877\\
12	-0.156592995634125\\
12.1	-0.103725515302021\\
12.2	0.0193312030110663\\
12.3	0.107064802543978\\
12.4	-0.0434657766871686\\
12.5	0.0390062329828555\\
12.6	0.166583193369898\\
12.7	0.14305511906841\\
12.8	0.130485774693332\\
12.9	0.0901441995735988\\
13	-0.0151396538336432\\
13.1	0.236457453666546\\
13.2	0.00375818826784338\\
13.3	0.0841683822566462\\
13.4	0.242858350695544\\
13.5	0.227734494097865\\
13.6	0.108138215697092\\
13.7	0.0189110855470917\\
13.8	0.0589195817475624\\
13.9	0.0965359180071182\\
14	-0.0607808676035776\\
14.1	-0.0971355459134453\\
14.2	0.0843401555619961\\
14.3	-0.0210880327662142\\
14.4	-0.124327286969645\\
14.5	0.0669953802704987\\
14.6	-0.146039503190516\\
14.7	-0.11050848778275\\
14.8	-0.164280974292707\\
14.9	-0.0952384862185908\\
15	-0.13244878686813\\
15.1	-0.26136665700234\\
15.2	-0.210234029656494\\
15.3	-0.23147706317234\\
15.4	-0.0760066681537637\\
15.5	0.0845435358170633\\
15.6	-0.0938641212954241\\
15.7	-0.159007758515003\\
15.8	0.0221454340907997\\
15.9	0.136517131290156\\
16	0.152281186970984\\
16.1	0.125080076231323\\
16.2	-0.026191909193886\\
16.3	0.0873722990238321\\
16.4	0.0238918647189726\\
16.5	0.0519355568335691\\
16.6	0.220490953130198\\
16.7	0.0896773840377011\\
16.8	0.0930050904225123\\
16.9	0.073230419952608\\
17	0.189602922267245\\
17.1	0.111441440293483\\
17.2	0.044720671007831\\
17.3	0.297028455641531\\
17.4	0.0417690262538512\\
17.5	0.15166316286044\\
17.6	0.095484511205781\\
17.7	0.100206714353766\\
17.8	-0.0831862125945941\\
17.9	-0.0388081901195016\\
18	-0.00685419522799301\\
18.1	-0.0540136215730776\\
18.2	-0.253525491072262\\
18.3	-0.155675836741322\\
18.4	-0.100315690947062\\
18.5	0.0362762427963024\\
18.6	-0.185458912331048\\
18.7	-0.029352044161341\\
18.8	-0.0943926104870743\\
18.9	0.0472994332306821\\
19	-0.191013684508108\\
19.1	-0.0503356252092248\\
19.2	-0.0410109186276183\\
19.3	-0.109976705588856\\
19.4	0.0646591897202008\\
19.5	-0.0574578263510847\\
19.6	-0.0395874555691342\\
19.7	-0.00637280539707354\\
19.8	-0.127717049957225\\
19.9	-0.0582613246198231\\
20	-0.0752594607747157\\
20.1	0.0557416491114109\\
20.2	-0.0201226603914241\\
20.3	0.144871939260155\\
20.4	0.207757954663116\\
20.5	-0.12663476457951\\
20.6	0.00476923081805912\\
20.7	-0.0958005150623892\\
20.8	0.226689485376074\\
20.9	0.0160338723170191\\
21	0.0707322501709223\\
21.1	-0.0136804023810897\\
21.2	0.0998006390637194\\
21.3	-0.207814108122245\\
21.4	-0.106194609285401\\
21.5	0.0601122126325383\\
21.6	0.0904379944203266\\
21.7	0.0293809627745547\\
21.8	0.131027920240983\\
21.9	0.0241558872631009\\
22	-0.0436014853222018\\
22.1	-0.103948208534635\\
22.2	-0.0113984834873911\\
22.3	-0.176824323822134\\
22.4	-0.156370809326959\\
22.5	-0.139231975948287\\
22.6	-0.0117448229981215\\
22.7	-0.0984762028554284\\
22.8	-0.183868446079864\\
22.9	-0.177409503893508\\
23	-0.0697000226852274\\
23.1	0.144931823103288\\
23.2	0.01481302466095\\
23.3	-0.0841388299973452\\
23.4	-0.0784151534091512\\
23.5	-0.00555565189034433\\
23.6	0.0720399570115124\\
23.7	-0.0305964110589134\\
23.8	-0.0910121767106794\\
23.9	-0.013215192631757\\
24	0.185529156476961\\
24.1	0.0446161989811591\\
24.2	-0.0187061354659214\\
24.3	0.0720097850107156\\
24.4	-0.0931230327815003\\
24.5	0.0692484600207006\\
24.6	0.0106083563347395\\
24.7	-0.097806500955521\\
24.8	-0.0911162924963653\\
24.9	-0.108893441101155\\
25	-0.150056981856737\\
25.1	-0.151516915568226\\
25.2	0.153434452147801\\
25.3	-0.0748587646100251\\
25.4	-0.00946526222044076\\
25.5	-0.162930993672006\\
25.6	-0.0245258173806905\\
25.7	-0.129379995958475\\
25.8	-0.123592063408834\\
25.9	0.0069991439186571\\
26	-0.188398156527578\\
26.1	-0.0277403804280639\\
26.2	0.194117638145239\\
26.3	-0.0499614579163933\\
26.4	0.116204678526593\\
26.5	-0.0936548543846003\\
26.6	0.032955507216954\\
26.7	0.127457326238274\\
26.8	0.180404818902324\\
26.9	-0.0117521082953292\\
27	0.106291955229318\\
27.1	0.0959677401687631\\
27.2	-0.0659259738794613\\
27.3	-0.0458432717337502\\
27.4	-0.0545373954641189\\
27.5	-0.112225644407711\\
27.6	0.074004162692327\\
27.7	-0.275072724604219\\
27.8	0.0059916902629849\\
27.9	-0.0296930996841369\\
28	0.0211034081607253\\
28.1	0.0963442009715613\\
28.2	-0.0729433488381186\\
28.3	-0.200488433888006\\
28.4	0.0671543642519993\\
28.5	0.0526672570393983\\
28.6	0.15499879095291\\
28.7	0.165790620691612\\
28.8	-0.0249734072879502\\
28.9	-0.179197627595453\\
29	0.0282056756087141\\
29.1	-0.12846440704508\\
29.2	0.0150793190115049\\
29.3	0.0125938757561173\\
29.4	-0.0716725062999313\\
29.5	0.0930594641680407\\
29.6	0.171773842868482\\
29.7	0.224377203956968\\
29.8	0.187199921839134\\
29.9	-0.00278686363217605\\
30	0.0907387211188824\\
30.1	0.0956656321821142\\
30.2	0.0458497551035432\\
30.3	-0.00124021818593292\\
30.4	-0.0293024262900758\\
30.5	0.135250024337982\\
30.6	0.0951564740037757\\
30.7	0.153653479915668\\
30.8	0.0282979173772953\\
30.9	0.000174462676436928\\
31	-0.0694336961360938\\
31.1	-0.167028561867439\\
31.2	0.179182153768655\\
31.3	-0.127553291827455\\
31.4	0.0924320842749135\\
31.5	-0.0621963770633636\\
31.6	-0.134023667701359\\
31.7	-0.26335644491533\\
31.8	-0.0412554905839261\\
31.9	0.0870030168613422\\
32	-0.099858911749424\\
32.1	0.0646243520369565\\
32.2	-0.0728717954180508\\
32.3	-0.00154426502166315\\
32.4	-0.0630227004677238\\
32.5	0.0782736838397881\\
32.6	0.0199081898911895\\
32.7	0.0514469312599604\\
32.8	0.169633095721603\\
32.9	-0.0192674648703428\\
33	0.130205009378368\\
33.1	0.00938353400229625\\
33.2	0.162881068359467\\
33.3	0.0714713172726722\\
33.4	0.112389939050864\\
33.5	0.00315126933196915\\
33.6	0.0511634512188521\\
33.7	0.303234366295091\\
33.8	0.12350891250587\\
33.9	0.0985318546619211\\
34	-0.0547185326107842\\
34.1	-0.230217222390685\\
34.2	-0.0559863984253789\\
34.3	-0.00247342655983111\\
34.4	0.0795664205268431\\
34.5	0.00856889059842676\\
34.6	-0.082576768486675\\
34.7	-0.27639152641558\\
34.8	-0.199906165915463\\
34.9	-0.0677379072993105\\
35	0.00187037766002479\\
35.1	-0.204311036643057\\
35.2	-0.0230479153005586\\
35.3	-0.185216076219857\\
35.4	0.0470889232504241\\
35.5	-0.0639290700463796\\
35.6	0.209175996051799\\
35.7	-0.171644869520422\\
35.8	-0.082544599464476\\
35.9	-0.128313298074255\\
36	0.139072234861729\\
36.1	-0.0423232926365413\\
36.2	0.312083785006353\\
36.3	0.181197763941729\\
36.4	0.248315980969763\\
36.5	0.251984005022435\\
36.6	0.0568917024278682\\
36.7	0.199416084053245\\
36.8	0.0539315461611261\\
36.9	0.206076244511921\\
37	0.023921384559931\\
37.1	0.106008552224009\\
37.2	0.13483043991745\\
37.3	0.240459022995348\\
37.4	-0.0130099155782253\\
37.5	-0.10789416024944\\
37.6	0.113584288060884\\
37.7	0.126140249182004\\
37.8	-0.0578068829447478\\
37.9	-0.0461004825187914\\
38	0.0651297112734418\\
38.1	-0.045639048103994\\
38.2	0.0247568981999364\\
38.3	-0.0809996977094174\\
38.4	-0.155941027433454\\
38.5	-0.213146772754277\\
38.6	-0.00374173061302385\\
38.7	-0.00522096474277636\\
38.8	-0.130340581601083\\
38.9	-0.00881358554978566\\
39	0.00648492320458598\\
39.1	-0.0653537740378924\\
39.2	0.209335415597256\\
39.3	0.100382830437807\\
39.4	-0.107347233090511\\
39.5	0.0157401492354347\\
39.6	0.133752336710436\\
39.7	-0.0738856861775687\\
39.8	0.0458325477592297\\
39.9	0.176920389965148\\
40	0.181210772220774\\
40.1	-0.0112809953961246\\
40.2	-0.0760113349209492\\
40.3	0.0903391179549004\\
40.4	0.199394192603226\\
40.5	0.077390147102226\\
40.6	-0.00166652851959619\\
40.7	0.138566485328345\\
40.8	0.102267674919277\\
40.9	-0.0111041639525267\\
41	0.0316731555905439\\
41.1	0.162428180943228\\
41.2	0.00261549673550261\\
41.3	0.0193548360387939\\
41.4	-0.0630666189763548\\
41.5	-0.202597689139747\\
41.6	-0.23444768546827\\
41.7	-0.00926856726385825\\
41.8	-0.0698340444976147\\
41.9	0.0273399913646235\\
42	-0.104850622356452\\
42.1	0.0510916936287302\\
42.2	0.0262794208611352\\
42.3	0.0326771289376444\\
42.4	0.122205038731039\\
42.5	-0.0674290603835376\\
42.6	0.0351701097098984\\
42.7	-0.00775535118398693\\
42.8	0.112467437097692\\
42.9	0.116920982060593\\
43	0.292082327976267\\
43.1	0.0274137554549014\\
43.2	0.0874406341358988\\
43.3	0.0682064335183467\\
43.4	0.0485384543386995\\
43.5	0.0702180616989137\\
43.6	0.131282454450601\\
43.7	0.319789480729581\\
43.8	0.196579209273122\\
43.9	-0.0714893839338333\\
44	0.32338609647872\\
44.1	0.054093935661699\\
44.2	0.0884431537706054\\
44.3	0.00980557230693777\\
44.4	0.0455148864914502\\
44.5	0.0188729008695232\\
44.6	-0.0612565966642686\\
44.7	-0.0952647154067366\\
44.8	0.00755423075431644\\
44.9	-0.0542992151024502\\
45	-0.237014476586832\\
45.1	-0.162545445763284\\
45.2	-0.0701136045193042\\
45.3	0.0974141902705876\\
45.4	0.0514453281699862\\
45.5	-0.0334988946283881\\
45.6	-0.168284474156943\\
45.7	0.0296971223676503\\
45.8	-0.099452218583256\\
45.9	-0.19241197079791\\
46	-0.187932034632616\\
46.1	0.0893808426312196\\
46.2	-0.0387124381711639\\
46.3	0.118044337994889\\
46.4	0.0937540187107331\\
46.5	0.10424055412562\\
46.6	0.102443778929065\\
46.7	0.0305145477084498\\
46.8	0.0998953925056331\\
46.9	0.276884776470159\\
47	0.267328582786956\\
47.1	0.000138877267170226\\
47.2	0.222949621228295\\
47.3	0.0158644823873134\\
47.4	0.170057905718324\\
47.5	0.0562329772037697\\
47.6	0.0860067663290086\\
47.7	-0.111364702676987\\
47.8	0.0929432204797857\\
47.9	-0.181361616672832\\
48	0.100916719942466\\
48.1	-0.148032273980069\\
48.2	-0.146509555207925\\
48.3	-0.0215575246188691\\
48.4	-0.221161474640673\\
48.5	-0.104166302764716\\
48.6	-0.100590170478784\\
48.7	-0.0909975290564898\\
48.8	0.0952673780019188\\
48.9	-0.227835708756994\\
49	-0.174811028563754\\
49.1	-0.0425869260350967\\
49.2	0.0868323609839768\\
49.3	0.0680312840103888\\
49.4	0.0581646334817368\\
49.5	0.0973983999113276\\
49.6	0.112317382640663\\
49.7	0.135791731737321\\
49.8	0.176932012073287\\
49.9	0.330473580566702\\
50	0.0592840451671096\\
50.1	0.0649628384563121\\
50.2	0.033835996448701\\
50.3	0.0912635230641819\\
50.4	0.0133459860601345\\
50.5	0.0346150093568823\\
50.6	-0.0392026961226264\\
50.7	0.0288955541879737\\
50.8	0.193976079000352\\
50.9	0.26851081487696\\
51	0.0809507799759382\\
51.1	-0.141326285265729\\
51.2	0.0324309028083409\\
51.3	0.0618773785659115\\
51.4	-0.0552239212018712\\
51.5	-0.0237403522689144\\
51.6	0.108607000298959\\
51.7	-0.240390357096668\\
51.8	-0.136267832473668\\
51.9	0.0941965713310685\\
52	-0.0204414006102906\\
52.1	-0.167972853280855\\
52.2	-0.170155673877061\\
52.3	0.0584033981392781\\
52.4	0.0599747330967212\\
52.5	-0.22702438868587\\
52.6	0.059408566940964\\
52.7	-0.12669743134769\\
52.8	-0.0916063688589436\\
52.9	0.0355689066155397\\
53	0.0982674069105133\\
53.1	0.0381645639541163\\
53.2	0.0861231388445652\\
53.3	0.0719826123964509\\
53.4	-0.069087088681228\\
53.5	0.125151618370568\\
53.6	-0.0455887644560287\\
53.7	-0.180426587524821\\
53.8	-0.00251448885216482\\
53.9	-0.0669161000819985\\
54	-0.0274884704085156\\
54.1	0.128989824990787\\
54.2	-0.183728086457545\\
54.3	0.0406721238348022\\
54.4	-0.0466450498920147\\
54.5	-0.0375409939951559\\
54.6	-0.138319622250332\\
54.7	0.0647006194863461\\
54.8	-0.0147863304763621\\
54.9	0.0982754849982179\\
55	-0.2542871574537\\
55.1	-0.176846967796391\\
55.2	-0.186581769251615\\
55.3	-0.121374624768845\\
55.4	-0.023637841000029\\
55.5	-0.106827233789939\\
55.6	-0.227300816709216\\
55.7	0.0229818789366348\\
55.8	-0.317168215042541\\
55.9	-0.0731055157355562\\
56	-0.0424852892941531\\
56.1	-0.171052294778168\\
56.2	0.0361353678184946\\
56.3	-0.127340661326618\\
56.4	-0.0903538475302888\\
56.5	0.146654491195375\\
56.6	-0.105868419381692\\
56.7	-0.0020357606430651\\
56.8	0.0652947076935824\\
56.9	-0.0402800904861823\\
57	0.129242528959017\\
57.1	0.0328924187976076\\
57.2	0.114546472681671\\
57.3	0.131967397172926\\
57.4	-0.119064496335502\\
57.5	0.00693237466214289\\
57.6	0.100239532899106\\
57.7	0.0441162525521791\\
57.8	-0.0226593447820408\\
57.9	0.0868237532283724\\
58	0.0363403113188344\\
58.1	0.137867299113512\\
58.2	0.17202069672517\\
58.3	0.031888096810492\\
58.4	0.142955377151495\\
58.5	-0.0733287118556651\\
58.6	0.00414164810995953\\
58.7	-0.202662032918542\\
58.8	-0.0670603782110223\\
58.9	0.0260489744322634\\
59	-0.101587333828641\\
59.1	-0.10350350097235\\
59.2	-0.112270775194249\\
59.3	-0.0209881612119177\\
59.4	-0.0176899603060525\\
59.5	-0.113981663965712\\
59.6	0.0714861083028721\\
59.7	-0.0162700930165526\\
59.8	-0.138183578019146\\
59.9	0.064312390256058\\
60	0.115661312958506\\
60.1	-0.11751048749315\\
60.2	-0.0645681022976408\\
60.3	-0.0882126831631611\\
60.4	0.103162748997842\\
60.5	-0.0581429276824573\\
60.6	0.124425918371315\\
60.7	0.0540508689398394\\
60.8	0.025861244029282\\
60.9	-0.0661249556066616\\
61	0.0671420730048775\\
61.1	-0.00946867814547441\\
61.2	-0.0854912810930733\\
61.3	0.0988229712179378\\
61.4	-0.00452277992081837\\
61.5	-0.193861174743521\\
61.6	0.0410542487441896\\
61.7	-0.00302165054628119\\
61.8	-0.14177989550446\\
61.9	-0.0984815400631622\\
62	-0.0806173072889812\\
62.1	0.042061554824311\\
62.2	-0.0905650130642564\\
62.3	-0.138299871080066\\
62.4	-0.0731576161753577\\
62.5	-0.0957060395356209\\
62.6	-0.0128671006068654\\
62.7	0.012115991939627\\
62.8	-0.0973641235252822\\
62.9	0.167514378256453\\
63	0.00150429978892089\\
63.1	-0.11352509725119\\
63.2	-0.0365230339914075\\
63.3	0.102001718088568\\
63.4	0.182571807171071\\
63.5	0.164991385608728\\
63.6	0.0714377916582433\\
63.7	0.192645539125537\\
63.8	-0.0546571548606497\\
63.9	0.176662306613676\\
64	0.095760652154871\\
64.1	0.220448022482275\\
64.2	0.042687244793712\\
64.3	0.0772051006016603\\
64.4	0.0997130948642095\\
64.5	0.136301694703478\\
64.6	0.031417114743325\\
64.7	0.131721279118211\\
64.8	-0.159167366217471\\
64.9	-0.0679366863906014\\
65	-0.0155696222726471\\
65.1	-0.0620296052157058\\
65.2	-0.23273257007767\\
65.3	0.0304722171322572\\
65.4	-0.0505788836764898\\
65.5	-0.0659945515013141\\
65.6	0.125448487267947\\
65.7	-0.186875078057841\\
65.8	-0.0654864352959038\\
65.9	-0.0764000930052576\\
66	-0.0869571719026528\\
66.1	0.0855015745270382\\
66.2	-0.02337456842348\\
66.3	0.0229411859617244\\
66.4	0.2429886428047\\
66.5	0.000287919647048974\\
66.6	0.00389331662888649\\
66.7	0.0766740175910477\\
66.8	0.00484713656881603\\
66.9	0.0201825768015644\\
67	0.188640450044844\\
67.1	0.124533756818745\\
67.2	0.300170602257996\\
67.3	-0.083757198027412\\
67.4	0.0336722917052322\\
67.5	0.0547410196728116\\
67.6	0.151012170564519\\
67.7	0.0447674973809077\\
67.8	-0.0353257334100769\\
67.9	0.0458463003565494\\
68	0.0817549427141228\\
68.1	-0.0353654593391403\\
68.2	0.00507581643088719\\
68.3	-0.149152126075565\\
68.4	0.0651560791402475\\
68.5	-0.179631883582967\\
68.6	-0.0854097475351709\\
68.7	-0.00194869633690604\\
68.8	-0.0106565040378789\\
68.9	-0.0826434285633471\\
69	0.0237591991567799\\
69.1	-0.263377760552484\\
69.2	-0.0831436116105564\\
69.3	0.1328415938267\\
69.4	0.0828461690247877\\
69.5	0.0115375660717168\\
69.6	0.139213986911158\\
69.7	-0.0711513816410758\\
69.8	0.0595658437955594\\
69.9	-0.152749734285142\\
70	0.0696036466200222\\
70.1	-0.0555618358189408\\
70.2	0.110644226042167\\
70.3	0.0693445876620779\\
70.4	0.00175863831118585\\
70.5	0.106444711878521\\
70.6	0.0917387278635245\\
70.7	0.106908243454547\\
70.8	0.0307001915167015\\
70.9	0.114928331171502\\
71	-0.0225173508418984\\
71.1	0.134999062466749\\
71.2	-0.200615597272591\\
71.3	-0.125605229251305\\
71.4	-0.156602051991824\\
71.5	-0.0138117737370379\\
71.6	-0.101464876789909\\
71.7	0.0701207432534616\\
71.8	-0.279082672845155\\
71.9	-0.104180094025562\\
72	-0.322736409253485\\
72.1	-0.0754622415483563\\
72.2	-0.125053608765879\\
72.3	-0.127037558139425\\
72.4	-0.0471116817317471\\
72.5	0.0376259766015956\\
72.6	0.00172483646124178\\
72.7	-0.00581511880657993\\
72.8	0.0166060828227144\\
72.9	0.0149699118945452\\
73	-0.214105670460943\\
73.1	0.0366881253946146\\
73.2	0.00661362969356341\\
73.3	0.0509886427364427\\
73.4	-0.0230765865957881\\
73.5	-0.0327798285103744\\
73.6	0.127792519341652\\
73.7	0.167236548865758\\
73.8	0.162847860357484\\
73.9	0.274728612164508\\
74	0.0188030583062761\\
74.1	0.0991048854840314\\
74.2	0.0886534014583707\\
74.3	0.0196143538567418\\
74.4	0.00390104025642546\\
74.5	-0.195867350429851\\
74.6	0.0981929243521416\\
74.7	-0.0664508667591588\\
74.8	0.183372497434618\\
74.9	0.134252642339295\\
75	0.0169090672012554\\
75.1	-0.159685253253605\\
75.2	-0.0205321040777297\\
75.3	0.150239507133516\\
75.4	-0.221733182418764\\
75.5	-0.073175089351287\\
75.6	-0.198573154431379\\
75.7	-0.0354738928885574\\
75.8	-0.222730902195575\\
75.9	-0.0679093094743857\\
76	-0.120297130111631\\
76.1	0.0303271497544864\\
76.2	0.0196816415661656\\
76.3	0.0669686384085715\\
76.4	0.108796869700456\\
76.5	-0.00184383290727669\\
76.6	0.124299370382707\\
76.7	0.167014899673295\\
76.8	0.0955951742618985\\
76.9	0.0333827412581781\\
77	0.0553214637615253\\
77.1	0.159445597059368\\
77.2	0.245888835898075\\
77.3	-0.0426640681297211\\
77.4	-0.0829672810423693\\
77.5	-0.107823099113821\\
77.6	-0.178686507286572\\
77.7	0.162421540023035\\
77.8	0.00393595691308814\\
77.9	-0.0537983458351267\\
78	-0.057320537582062\\
78.1	-0.0384164079934722\\
78.2	-0.0525206238325516\\
78.3	-0.175288394288637\\
78.4	-0.0792483057378038\\
78.5	-0.170460242929667\\
78.6	0.231020181231132\\
78.7	0.0144133400440274\\
78.8	-0.253094674425717\\
78.9	-0.162400032697353\\
79	-0.111581366588362\\
79.1	-0.15826013003197\\
79.2	-0.02393620981606\\
79.3	-0.289136004363815\\
79.4	0.163267478077584\\
79.5	-0.0965627367252293\\
79.6	0.00919607869431652\\
79.7	0.0292572831859687\\
79.8	0.0275273061477156\\
79.9	0.221137456810393\\
80	0.0838061134507684\\
80.1	0.0627963583032926\\
80.2	0.180483871912254\\
80.3	0.0850186343001396\\
80.4	0.13193918254004\\
80.5	0.168752382776743\\
80.6	0.0771917918117456\\
80.7	0.172429274038199\\
80.8	0.225515060536676\\
80.9	0.068764795356782\\
81	0.125531814541776\\
81.1	0.111679205204826\\
81.2	0.152232206678796\\
81.3	0.123975803162048\\
81.4	-0.0812583010042239\\
81.5	-0.183831802836161\\
81.6	0.0126620885178911\\
81.7	-0.104682038644204\\
81.8	-0.0739910102016247\\
81.9	-0.166320258941633\\
82	-0.0668397088187514\\
82.1	-0.0701913503206483\\
82.2	-0.216800579448525\\
82.3	0.00136402385535231\\
82.4	-0.155479219706341\\
82.5	0.0392829530618762\\
82.6	-0.199520981979242\\
82.7	0.0286434065906951\\
82.8	-0.033875563743297\\
82.9	0.0273201871875119\\
83	0.0444353601061846\\
83.1	0.0771049123502469\\
83.2	-0.124477541840924\\
83.3	0.104566759819978\\
83.4	-0.0958183033664635\\
83.5	0.165912125380845\\
83.6	-0.101352881824282\\
83.7	0.00927301285879972\\
83.8	0.0326864751138676\\
83.9	0.0417952775495545\\
84	0.094603192776194\\
84.1	0.0716344284183816\\
84.2	0.178336262342717\\
84.3	-0.0917937724664621\\
84.4	0.0436504200972668\\
84.5	0.107619239648599\\
84.6	0.0100181756695503\\
84.7	0.055134905227052\\
84.8	-0.0811283973596725\\
84.9	-0.0216728117140722\\
85	-0.109961970526977\\
85.1	-0.0385297071542502\\
85.2	-0.0352212145600149\\
85.3	-0.178488288132911\\
85.4	-0.182357374086967\\
85.5	-0.145092814321304\\
85.6	-0.090860968679435\\
85.7	-0.0402593851308729\\
85.8	-0.071531356785682\\
85.9	-0.0437767436364705\\
86	0.0481248433619163\\
86.1	-0.0505354863556056\\
86.2	-0.0337716837037382\\
86.3	-0.143043696251221\\
86.4	-0.0429951877321416\\
86.5	0.0628683654309986\\
86.6	0.0670247078295827\\
86.7	-0.108870387607026\\
86.8	0.0982347887250109\\
86.9	0.109393552024125\\
87	-0.0595494292923077\\
87.1	0.024204960808348\\
87.2	-0.122717832663827\\
87.3	0.204342156209959\\
87.4	-0.000589278141353837\\
87.5	0.00453686177575131\\
87.6	-0.0609748551973595\\
87.7	0.00912808145514392\\
87.8	-0.0421799091741131\\
87.9	-0.179857413354725\\
88	-0.00995816715542051\\
88.1	0.116550180705255\\
88.2	-0.00712513442225071\\
88.3	0.239813787826353\\
88.4	0.0471016784066476\\
88.5	-0.0313705327449373\\
88.6	-0.0663491730848289\\
88.7	-0.0116633983241816\\
88.8	-0.161331180321439\\
88.9	-0.029188046299496\\
89	0.072707083011349\\
89.1	-0.0480785367094188\\
89.2	-0.0583861974032955\\
89.3	-0.0419203196585148\\
89.4	-0.0109247304207459\\
89.5	-0.171180146050273\\
89.6	-0.0388956812846435\\
89.7	0.0511543346736522\\
89.8	-0.117451544649521\\
89.9	0.0992410753994479\\
90	0.0444263118484012\\
90.1	0.0121773423579417\\
90.2	-0.0772976664982973\\
90.3	0.0557635150757131\\
90.4	-0.0186912870163414\\
90.5	-0.0247748060842461\\
90.6	0.0318045228618085\\
90.7	-0.0222410156068363\\
90.8	0.0693137762462986\\
90.9	0.0745846115186038\\
91	-0.0595724121971045\\
91.1	-0.121872569656225\\
91.2	-0.0348584526099073\\
91.3	0.0780552787024978\\
91.4	0.0744583707611353\\
91.5	-0.024270160535576\\
91.6	0.0026128008845137\\
91.7	-0.0312251498497253\\
91.8	0.061383825739217\\
91.9	-0.0665547476475991\\
92	-0.0899277855602767\\
92.1	0.0724066701975004\\
92.2	0.0726033525168627\\
92.3	0.017615957854615\\
92.4	-0.169874026365009\\
92.5	-0.0162278038937833\\
92.6	-0.0809074241981161\\
92.7	-0.00374197620945339\\
92.8	0.119665487954215\\
92.9	-0.0460203476439578\\
93	0.200558547234098\\
93.1	0.0569401671623831\\
93.2	0.0745388625232973\\
93.3	0.144095046421065\\
93.4	-0.0347959366879114\\
93.5	0.0408969052455542\\
93.6	0.0854183276755999\\
93.7	-0.0966330807470192\\
93.8	0.0502035383939851\\
93.9	0.147802379196847\\
94	0.0612182863345532\\
94.1	0.0790959086201339\\
94.2	0.10011330393395\\
94.3	0.116485253457142\\
94.4	0.0278784910128577\\
94.5	0.0105576782640263\\
94.6	-0.0631029536699636\\
94.7	-0.0714136632266613\\
94.8	-0.130400645302181\\
94.9	-0.0868598752424701\\
95	-0.029031451907046\\
95.1	-0.198444987019125\\
95.2	-0.137988135027186\\
95.3	-0.147833427295971\\
95.4	-0.134258909967206\\
95.5	-0.112836529761624\\
95.6	-0.0509824540796149\\
95.7	-0.0895816493498184\\
95.8	0.0785346739550669\\
95.9	-0.0208318744635181\\
96	-0.115548995264926\\
96.1	-0.124873087022962\\
96.2	0.0130806983310936\\
96.3	0.0196106439305288\\
96.4	-0.0953181518871987\\
96.5	-0.116388208707351\\
96.6	0.051051337285087\\
96.7	0.209075231015089\\
96.8	0.10647645321645\\
96.9	0.135481227268903\\
97	0.09451765784228\\
97.1	-0.0379330454596545\\
97.2	0.214450788500226\\
97.3	0.0805977896786017\\
97.4	0.00950624444260913\\
97.5	0.087923776975337\\
97.6	0.169126987074706\\
97.7	-0.213980281745375\\
97.8	0.169787987941447\\
97.9	0.212790430591728\\
98	0.0389700912516194\\
98.1	0.1153059652284\\
98.2	-0.10830899860938\\
98.3	-0.0999663197558618\\
98.4	0.0201861418982312\\
98.5	-0.0136330954542278\\
98.6	0.0136116329560208\\
98.7	-0.218440064881337\\
98.8	-0.0120402276614939\\
98.9	-0.118471151970429\\
99	-0.114991578393149\\
99.1	-0.0786780560333901\\
99.2	-0.225209590025628\\
99.3	-0.142435217589165\\
99.4	0.0982463140544126\\
99.5	-0.121117363927699\\
99.6	-0.104893862067621\\
99.7	-0.0509409883914199\\
99.8	-0.163220080992689\\
99.9	0.0370867839759599\\
100	0.0193081767982295\\
};
\end{axis}
\end{tikzpicture}%

%% file: discussion.tex
\section{Discussion}
\label{sec:discussion}
%
%
Probabilistic modeling is about uncertain representations of data and knowledge using probability distributions and how to actually compute these representations by inference and learning algorithms. We have in this tutorial explained a solution to the problem of computing the posterior distribution $p(\theta\mid y_{1:\T})$ of the unknown parameters $\theta$ in a nonlinear state-space model~\eqref{eq:SSM} using approximate solutions that converge to that posterior distribution as we make use of more computational power. This solution involved the use of a particle filter inside a standard Metropolis--Hastings algorithm, where the particle filter was used to compute non-negative and unbiased estimates of the likelihood which in turn enabled the computation of the relevant acceptance probabilities. There are fairly general software implementations available via the modeling language LibBi \cite{Murray:2015} which was used to implement the example.
%
We will end this tutorial by briefly reflecting upon some recent developments on probabilistic \emph{representations} and how to carry out the \emph{computations} required to learn and make use of these representations.

%
The development of new probabilistic models describing nonlinear dynamical phenomena is a highly active and exciting area of research right now. One of the key lessons from modern machine learning is that flexible models often provide the best results \citep{Ghahramani:2015}. The two dominating approaches for creating flexible models are Bayesian non-parametrics \citep{Ghahramani:2013} and deep learning \citep{LeCunBH:2015}, both of which can be combined with the state-space model to create flexible and useful models. The most popular Bayesian non-parametric model is arguably the Gaussian process \citep{RasmussenW:2006} which offers a probabilistic distribution over functions. The Gaussian process construction thus offers a rather natural probabilistic model for the nonlinear functions $f(\cdot)$ and $g(\cdot)$ in~\eqref{eq:SSM}. Such a construction was developed in~\cite{FrigolaLSR:2013} and the required learning algorithms were recently enhanced in \cite{SvenssonS:2017} based on new model approximations. When using flexible models it is important to have a principled way of trading-off their flexibility and the actual fit to data. An emerging way of developing flexible models is provided by probabilistic programming where probabilistic models are represented using computer programs, which was in fact used to implement and solve the example we gave in this tutorial.

%
The development of new and more capable approximations based on sequential Monte Carlo methods is progressing rapidly.
As a first clear trend we mention algorithms scaling to higher-dimensional models, where many of the standard algorithms struggle. Relevant developments in this direction involve localization strategies  \citep{RebeschinivH:2015}, the so-called location particle smoother \citep{BriggsDM:2013} and the discussion in \cite{DjuricB:2013}. Furthermore, the nested SMC method \citep{NaessethLS:2015} allows us to \emph{exactly approximate} the locally optimal proposal, and significantly extend the class of models for which we can perform efficient inference using SMC. This development also opens up approximate inference in nonlinear spatio-temporal state-space models \citep{Wikle:2015,CressieW:2011}.

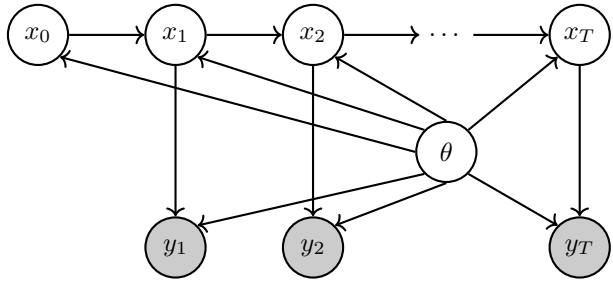
\begin{figure}[ht]
	\centering
	\begin{tikzpicture}
	[place/.style={circle,draw=black!,thick,inner sep=0pt,minimum size=8mm}]
	\node[place] (x0) {$x_0$};
	\node[place] (x1) [right=of x0] {$x_1$};
	\node[place] (x2) [right=of x1] {$x_2$};
	\node[] (dots) [right = of x2] {$\dots$};
	\node[place] (xN) [right = of dots] {$x_{\T}$};
	\node[place, fill=black!20] (y1) [below = of x1,yshift=-1cm] {$y_1$};
	\node[place, fill=black!20] (y2) [below = of x2,yshift=-1cm] {$y_2$};
	\node[place, fill=black!20] (yN) [below = of xN,yshift=-1cm] {$y_{\T}$};
	\node[place] (th) [below = of dots] {$\theta$};
	\draw[thick] [->] (x0.east) -- (x1.west);
	\draw[thick] [->] (x1.east) -- (x2.west);
	\draw[thick] [->] (x1.south) -- (y1.north);
	\draw[thick] [->] (x2.south) -- (y2.north);
	\draw[thick] [->] (x2.east) -- (dots.west);
	\draw[thick] [->] (dots.east) -- (xN.west);
	\draw[thick] [->] (xN.south) -- (yN.north);
	\draw[thick] [->] (th.west) -- (x0.south east);
	\draw[thick] [->] (th.north west) -- (x1.south east);
	\draw[thick] [->] (th.north) -- (x2.south east);
	\draw[thick] [->] (th.south west) -- (y1.north east);
	\draw[thick] [->] (th.south) -- (y2.north east);
	\draw[thick] [->] (th.north east) -- (xN.south west);
	\draw[thick] [->] (th.south east) -- (yN.north west);
	\end{tikzpicture}
	\caption{Probabilistic graphical model representing the \ssm introduced in~\eqref{eq:SSMpdf} for $\T$ measurements.}
	\label{fig:PGM_SMC}
\end{figure}

A probabilistic graphical model \citep{Jordan:2004,WainwrightJ:2008,KollerF:2009} makes use of a graph $\mathcal{G} = (\mathcal{V}, \mathcal{E})$ to represent the relationships between the random variables used in the model. Each random variable is represented as a vertex/node in the graph and the set of all vertices is denoted by $\mathcal{V}$. The set of edges $\mathcal{E}$ contain elements $(i, j)\in \mathcal{E}$ connecting the two nodes $(i, j)\in \mathcal{V} \times \mathcal{V}$. As an example we provide the directed probabilistic graphical model of the \ssm~\eqref{eq:SSMpdf} in Figure~\ref{fig:PGM_SMC}. From this figure it is clear that the \ssm is a very specific instance of a probabilistic graphical model,
which leads to the rather natural question of whether we can also make use of SMC for inference and learning in more general graphical models. The answer to this question is yes and one key is to introduce a \emph{sequential decomposition} of the graphical model. This sequential decomposition can then be exploited by SMC, since SMC is in fact applicable to any problem that amounts to learning a sequence of probability distributions defined on a sequence of spaces of increasing dimension. See e.g.,  \cite{NaessethLS:2014,BeskosCJKZ:2017stpf,Lindsten:2017DC-SMC} for developments in this direction. For a tutorial style introduction to the use of SMC for inference in probabilistic graphical models we refer to \citep{DoucetL:2017}. This ends our discussion on the second trend, namely the use of SMC type algorithms for inference in more general probabilistic models.

The third trend is the composition of two or more existing algorithms
into new and more capable algorithms. We have in this tutorial
presented the particle Metropolis--Hastings algorithm---which is one
instantiation of this trend---corresponding to a particular
combination of the particle filter and the Metropolis--Hastings
algorithm. It belongs to the growing family of exact approximation
algorithms.
The particle Gibbs sampler~\citep{AndrieuDH:2010} makes use of the
so-called conditional particle filter to generate samples from the
posterior distribution. The SMC$^2$ sampler which was independently
derived by \cite{ChopinJP:2013smc2,FulopL:2013} is similar in spirit
to the particle Metropolis--Hastings algorithm in that it makes use of
a particle filter within another sampling algorithm. More specifically
it makes use of an SMC sampler \citep{DelMoralDJ:2006} over parameters $\theta$ that in turn
has many internal particle filters over the state $x$. The particle filter can also be composed with another particle filter to produce interesting and capable algorithms. The idea of coupling two particle filters is developed in~\cite{JacobLS:2017} and the resulting construction can for example be used to build a competitive solution to the nonlinear state smoothing problems.

%% file: code.tex
\section{Implementation in LibBi}
\label{sec:code}
The numerical example was implemented using LibBi (\cite{Murray:2015},
\url{www.libbi.org}). The code needed to reproduce the examples will
be available via the first authors homepage. Apart from simulating
data, the LibBi code for implementing the model~\eqref{eq:ex:model}
and run the PMH learning algorithm is provided below.

First, the following content is placed in a file called \verb|Damper.bi|. 
\lstset{literate={~}{$\sim$}{1},columns=fixed,basewidth=0.5em,tabsize=2,basicstyle=\footnotesize\ttfamily,breaklines=true}

\begin{footnotesize}
\begin{lstlisting}
model Damper {
	/* Specifying all variables in the problem */
	const T_s = .1, m = 2           /* Known constants */
	param f_c, c_0, k, p            /* Unknown variables */
	noise v                         /* Process noise */
	state s, sdot, s_old, sdot_old  /* State variables */
	obs y

	sub parameter {
		/* Specifying all priors for the unknown parameters  */
		f_c ~ gamma(2, 0.01)
		c_0 ~ gamma(2, 1)
		k ~ gamma(4, 0.3)
		p ~ uniform(0, 1)
	}
	
	sub proposal_parameter {
		/* Specifying the random walk proposal used in PMH */
		f_c ~ truncated_gaussian(f_c, 1.0e-3, 0.0)
		c_0 ~ truncated_gaussian(c_0, 1.0e-2, 0.0, 1.0)
		k ~ truncated_gaussian(k, 1.0e-2, 0.0)
		p ~ truncated_gaussian(p, 1.0e-2, 0.0, 1.0)
	}

	sub initial {
		/* Specifying the inital state variables */
		s <- 0.5
		sdot <- 0.0
	}

	sub transition(delta = 1) {
		/* Describe how the states in the model evolves at each time step. The variables s_old and sdot_old is only for temprary storage of the previous state values. */
		v ~ gaussian(0.0, 1.0e-2)
		
		s_old <- s
		sdot_old <- sdot
		
		s <- s_old + T_s*(sdot_old)
		sdot <- sdot_old + (T_s/m)*(-f_c*((sdot_old < 0) ? -1 : 1) - c_0*sdot_old - k*((s_old < 0) ? -1 : 1)*pow(abs(s_old), p)) + v
	}

	sub observation {
		/* Describe how the measurements relates to the states */
		y ~ gaussian(s, 1.0e-1)
	}
}
\end{lstlisting}
\end{footnotesize}

Once the file \verb|Damper.bi| is in place, the following command can be run from the command line:

\begin{footnotesize}
\begin{lstlisting}
libbi sample --target posterior --model-file Damper.bi --nsamples 10000 --nparticles 256 --end-time 1000 --sampler mh --obs-file data/obs.nc --output-file results/posterior.nc
\end{lstlisting}
\end{footnotesize}

which will produce $10\thinspace 000$ samples from
$p(\theta\mid y_{1:1000})$, where $y_{1:1000}$ is found in the file
\verb|data/obs.nc|. The samples are generated by the PMH algorithm and
recorded in the file \verb|results/posterior.nc| which can be loaded
into, e.g., Python, Matlab or R for further processing.